\documentclass[12pt]{article}
\usepackage{graphicx}
\usepackage{epsfig}
\def\be{\begin{equation}} \def\ee{\end{equation}} \def\bea{\begin{eqnarray}}
\def\eea{\end{eqnarray}} \def\nnb{\nonumber}

\begin{document}
\hfill{LA-UR-08-1461}

\begin{center}
{\LARGE Parity-violating nucleon-nucleon interaction \\
\vskip 0.2cm

from different approaches}
\vskip 1cm

\renewcommand{\thefootnote}{\fnsymbol{footnote}}
\begin{large}
B.  Desplanques$^{1}$\footnote{{\it E-mail address:}  desplanq@lpsc.in2p3.fr}, 
C. H. Hyun$^{2}$\footnote{{\it E-mail address:}  hch@color.skku.ac.kr; 
now at Department of Physics Education, Daegu University, Gyeongsan 712-714,
Korea}, 
S. Ando$^{2}$\footnote{{\it E-mail address:} Shung-ichi.Ando@manchester.ac.uk;
now at Theoretical Physics Group, School of Physics and Astronomy, 
The Manchester University, Manchester M13 9PL, UK },
C.-P. Liu$^{3}$\footnote{{\it E-mail address:} cliu38@wisc.edu;
now at Department of Physics, University of Wisconsin-Madison,
1150 University Avenue, Madison, WI 53706-1390 }
\\
$^{1}$LPSC, Universit\'e Joseph Fourier Grenoble 1, CNRS/IN2P3, INPG, \\
  F-38026 Grenoble Cedex, France \\
$^{2}$Department of Physics and Institute of Basic Science,\\
Sungkyunkwan University, Suwon 440-746, Korea \\
$^{3}$T-16, Theoretical Division, Los Alamos National Laboratory, \\Los Alamos, NM
87545, USA

\vskip 0.5cm

March 14, 2008
\end{large}
\end{center}



\vskip 1.0cm


\begin{abstract}
\small{
\noindent
Two-pion exchange parity-violating nucleon-nucleon interactions 
from recent effective field theories and earlier fully covariant approaches 
are investigated. The potentials are compared with the idea to obtain 
better insight on the role of low-energy constants appearing 
in the effective field theory approach and the  convergence of this one 
in terms of a perturbative series. The results 
are illustrated by considering the longitudinal asymmetry 
of polarized protons scattering off protons, $\vec{p}+p \rightarrow p+p$, 
and the asymmetry of the photon emission in radiative capture 
of polarized neutrons by protons, $\vec{n}+p \rightarrow d+\gamma$. 
}
\end{abstract} 
\renewcommand{\thefootnote}{\arabic{footnote}}
\setcounter{footnote}{0}
\section{Introduction}
Effective field theory (EFT) underlies most recent developments in the
domain of the nucleon-nucleon ($NN$) strong interaction 
\cite{ork_prc96,egm_npa00,entem_prc68,epel_05}. The approach is mainly 
motivated by the fact that a large part of the short-range interaction 
is essentially unknown. Its detailed description  may not be really
relevant at low energy and a schematic one, represented 
by contact interactions with low-energy constants (LEC's), 
could be sufficient. Moreover, such an approach 
could account for important properties in relation to QCD dynamics,   
i.e. chiral symmetry. Implementing these properties
can be done with the chiral perturbation theory \cite{weinberg}. 
One is thus led to distinguish contributions at different orders.
Beyond the one-pion exchange (OPE), which appears at leading order (LO), 
two-pion exchange (TPE), which is relevant at next-to-next-to-leading order 
(NNLO), has been considered.\footnote{There are different conventions to denote 
orders.  We use the one in agreement with what has been used 
in the parity-violating case.} 
Higher order terms are also considered, contributing to a successful 
description of the strong interaction.

Naturally, the EFT approach has been applied to the weak, 
parity-violating (PV), $NN$ interaction \cite{zhu_npa05}, 
superposing on earlier phenomenological works in the 70's 
\cite{danilov_sjnp72, missimer_prc76,desplanques_npa79} 
with a systematic perturbation scheme in terms  
of an expansion  parameter characterizing the theory. 
Thus, a one-pion-exchange contribution appears at LO 
while the two-pion-exchange contribution is part of those at NNLO. 
It was claimed that effects from the two-pion-exchange contribution 
could be potentially large. Estimates have been made for various observables  
\cite{hyun_plb07, liu_xxx06} and, while they do not contradict 
the above expectation, they have evidenced a large range 
of uncertainty \cite{hyun_plb07}. This  points out to the role 
of a contact term present in the operators, 
which has to be completed in any case by a LEC contribution 
and cannot be therefore considered as physically relevant by itself. 
The PV TPE contribution was considered in the 70's in several works starting 
from a covariant formalism and based on Feynman diagrams \cite{pign_plb71} 
or dispersion relations \cite{desp_plb72,pirner_plb73,chem_npa74}.
Originally, these works were motivated by the expectation that the TPE could
play a role  in the PV case as important as  in the strong interaction one, 
what was actually disproved by the studies.
A similar motivation 
has recently been addressed within the EFT approach with some attention 
to the contribution involving the $\Delta$ excitation \cite{kaiser_prc07}
(see also Refs. \cite{liu_arxiv07,niskanen_arxiv07} with this last respect).
Interestingly, the TPE contribution in various processes turned out to be 
well determined but rather small \cite{desp_plb72,chem_npa74,desp_npa75} 
and, in particular,  unessential  in comparison with other uncertainties 
(PV couplings constants, nuclear-structure description, {\it etc}). 
This last feature largely explains its omission in later works.
In view of different conclusions, we believe that a comparison 
of both the recent and earlier works is useful. Some preliminary results 
were presented in Ref. \cite{hyun_chiral07}.

The above studies could be relevant because, 
contrary to the strong interaction, it is not
possible to determine at present the LEC's due to the lack 
of sufficient and accurate enough experimental data. 
On the one hand, they could tell us about the role of contact interactions 
in making the two approaches as close as possible. 
The contributions of these contact interactions can be ascribed 
to the LEC or the ``finite" range part, depending on the subtraction scheme.
On the other hand, they could provide information on the convergence 
of the perturbative expansion of the potential in the EFT approach, 
which is limited to NNLO so far. 
In the field of the strong $NN$ interaction \cite{higa06} 
or weak semi-leptonic interactions \cite{ando06}, 
there are hints for non-negligible corrections.

The plan of this paper is as follows. In the second section, 
results relative to the TPE contribution from a covariant approach 
are reminded (isovector part). All components of this interaction 
and their convergence properties are in particular discussed. 
The third section is devoted to results in the EFT approach, 
completed by those obtained from the contribution of time-ordered 
diagrams as a check. In the fourth section, we examine 
similarities and differences between results of these approaches 
and those from an expansion of the covariant one 
at the lowest non-zero order in the inverse of the nucleon mass, $1/M$.
A numerical comparison concerning a few aspects 
of potentials so obtained is given in the fifth section. 
Estimates of the effects in two selected processes, 
proton-proton scattering and radiative neutron-proton capture 
at thermal energy, are finally given in the sixth section. 
These two processes allow one to illustrate the two types 
of PV effects that are expected from the TPE interaction at low energy. 
The seventh section contains the conclusion. This is completed by appendices
concerning the removing of the iterated one-pion exchange and the EFT approach.


\section{Two-pion exchange from a covariant formalism}
\label{COV}
\begin{figure}[htb]
\begin{center}
\mbox{ \epsfig{ file=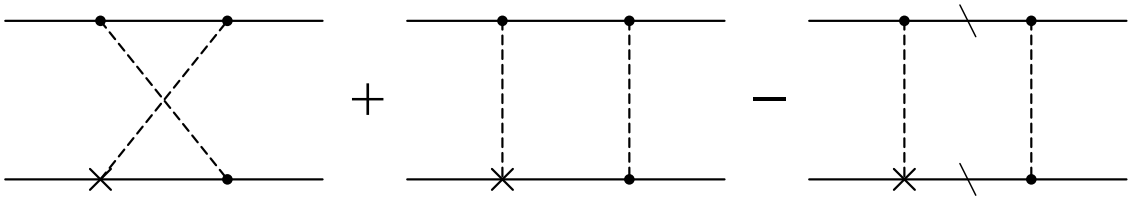, width=15cm}}
\end{center}
\caption{Two-pion exchange in the covariant approach:
These diagrams represent the contributions of the crossed box, 
the non-crossed box and the iterated OPE that has to be subtracted 
from the previous one. The continuous line represents a baryon 
and the dashed one a pion. The contributions with nucleon and nucleon resonances 
in the intermediate state have been considered in the literature. 
Only the first one is retained here but the role of the other ones 
will also be mentioned. The PV meson-nucleon vertex is marked 
with a cross, $\times$. The last diagram on the right represents 
the iterated one-pion exchange, the back slash indicating 
that the corresponding nucleon is on-mass shell. Further diagrams 
with a different order of the parity-violating and parity-conserving 
meson-nucleon vertices on the same nucleon line, not shown here, 
are also considered. 
\label{fig:covar}}
\end{figure} 
We here consider the isovector PV TPE contribution to the $NN$ interaction 
obtained from a fully covariant formalism.
It is induced by an elementary PV pion-nucleon coupling, 
most often denoted by $h^1_{\pi}$. As this coupling also determines the strength 
of the  PV OPE interaction, we give the corresponding expression 
in terms of both the momentum transfer, $\vec{q}$, 
and the relative momenta, $\vec{p}$ and $\vec{p}\,'$: 
\begin{eqnarray}
V_{\pi}(q)= \frac{i\,g_{\pi NN}\; h^1_{\pi}}{2\sqrt{2}\,M}\;
(\tau_1 \! \times \! \tau_2)^z\;
\frac{(\vec{\sigma}_1\!+\!\vec{\sigma}_2) \! \cdot \!
 (\vec{p}\,' \!  - \! \vec{p}\,)  }{m^2_{\pi}+q^2}  
 = - \frac{i\,g_{\pi NN}\; h^1_{\pi}}{2\sqrt{2}\,M}\;
(\tau_1 \! \times \! \tau_2)^z\;
\frac{(\vec{\sigma}_1\!+\!\vec{\sigma}_2) \! \cdot \!
\vec{q}}{m^2_{\pi}+q^2} 
\, .
\label{eq:onepi}
\end{eqnarray}
Due to possible ambiguity, we further specify the notations for momenta:
\begin{eqnarray}
&&\vec{p}=\frac{1}{2 }\,(\vec{p}_1-\vec{p}_2),\;\;
\vec{p}\,'=\frac{1}{2 }\,(\vec{p}\,'_1-\vec{p}\,'_2) \, ,
\nonumber \\
&&\vec{q}=(\vec{p}_1-\vec{p}\,'_1)= -(\vec{p}_2-\vec{p}\,'_2)
=(\vec{p}-\vec{p}\,')\, ,
\label{eq:defmts}
\end{eqnarray}
where the primed and non-primed  momenta  refer to  those of particles 
appearing respectively in the bra and ket states. 
In configuration space, the interaction thus recovers its standard form:
\begin{eqnarray}
V_{\pi}(r) &=& \frac{i\,g_{\pi NN}\; h^1_{\pi}}{2\sqrt{2}\,M}\;
(\tau_1 \times \tau_2)^z\;\;
(\vec{\sigma}_1\!+\!\vec{\sigma}_2)\!\cdot
\Big[\vec{p},\; \frac{e^{-m_{\pi}r}}{4\pi \,r}\Big] 
\nonumber \\
&=& -\frac{\,g_{\pi NN}\; h^1_{\pi}}{2\sqrt{2}\,M}\;
(\tau_1 \times \tau_2)^z\;\;
(\vec{\sigma}_1\!+\!\vec{\sigma}_2)\!\cdot\! (\vec{r_1}\!-\!\vec{r_2}) \;
\frac{e^{-m_{\pi}r}(1\!+\!m_{\pi}r)}{4\pi \,r^3} \, ,
\label{eq:onepir}
\end{eqnarray}
with $r=|\vec{r_1}-\vec{r_2}|$.
First studies of the isovector PV TPE contribution were made in the early 70's
\cite{pign_plb71,desp_plb72,pirner_plb73,chem_npa74}.
For a part, they were motivated by the underestimation of some 
observed PV effects using the standard PV $NN$ interaction 
available at that time. 
Similarly to the strong-interaction case, where the TPE contribution 
is quantitatively more important than the OPE one, it was believed that the  
TPE contribution could also play an essential role in the PV case.

Various studies roughly agree with each other, 
after correcting mistakes in some cases \cite{pign_plb71,pirner_plb73}. 
Differences involve in particular the formalism 
(calculations  using Feynman diagrams or dispersion relations), 
non-relativistic approximations in external nucleon lines, 
or the removing of the OPE-iterated contribution in the box diagram. 
The choice of the Green's function in the last ingredient was
rather unimportant due to the introduction of cutoffs in applications.  
It could however be important in unrestricted calculations. The point 
is of relevance with respect to a comment made in Ref. \cite{zhu_npa05}
about the absence of convergence in earlier calculations. 
It will be discussed in more detail below when expressions 
for the TPE contribution are given.

The crossed and non-crossed box diagrams that enter the isovector 
TPE contribution of interest here are represented in Fig. \ref{fig:covar}, 
where the intermediate hadron on the upper line can be a nucleon 
as well as a resonance ($\Delta(1231), \;N^*(1440),\; N^*(1518) $).
For our purpose and also for simplicity, we only retain the nucleon.
Altogether, the corresponding contribution to the isovector interaction 
involves six different terms.  
Following for a part the notations of Ref. \cite{chem_npa74}, it can be written 
quite generally in momentum space as:
\begin{eqnarray}
&&V(\vec{p}\,', \vec{p})=V_{44}+V_{34}+V_{56}+V_{75}+V_{66}+V_{85}
\nonumber \\
&&\hspace*{3mm}=
i\,(\tau_1+\tau_2)^z\;\;(\vec{\sigma}_1 \times \vec{\sigma}_2)\!\cdot\!
(\vec{p}\,'- \vec{p})\; v_{44}(q,\cdots)
\nonumber \\
&&\hspace*{6mm}+\,(\tau_1+\tau_2)^z\;\;(\vec{\sigma}_1 - \vec{\sigma}_2)\!\cdot\!
(\vec{p}\,'+ \vec{p})\;  v_{34}(q,\cdots)
\nonumber \\
&&\hspace*{6mm}+i\,(\tau_1\times\tau_2)^z\;\;(\vec{\sigma}_1 + \vec{\sigma}_2)\!\cdot\!
 (\vec{p}\,'- \vec{p})\; v_{56}(q,\cdots)
\nonumber \\
&&\hspace*{6mm}+\,(\tau_1-\tau_2)^z\;\;(\vec{\sigma}_1 + \vec{\sigma}_2)\!\cdot\!
(\vec{p}\,'+ \vec{p})\;  v_{75}(q,\cdots)
\nonumber \\
&&\hspace*{6mm}+\,(\tau_1 \!\times\! \tau_2)^z\;\;
\Big ( \vec{\sigma}_1 \!\cdot\! \vec{q}\;
\vec{\sigma}_2\!\cdot\!(\vec{p}\,'\!+\! \vec{p})\!\times\!\vec{q} +
(\vec{\sigma}_1 \leftrightarrow   \vec{\sigma}_2)
\Big)\;v_{66}(q,\cdots)
\nonumber \\
&&\hspace*{6mm}-i\,(\tau_1\!-\!\tau_2)^z\;\;
\Big (\vec{\sigma}_1\!\cdot\!(\vec{p}\,'\!+\! \vec{p})\; 
\vec{\sigma}_2\!\cdot\!(\vec{p}\,'\!+\! \vec{p})\!\times\!\vec{q} +
(\vec{\sigma}_1 \leftrightarrow   \vec{\sigma}_2)
\Big)\;v_{85}(q,\cdots)\,.
\label{eq:twopi}
\end{eqnarray}
Functions $v_{ij}(q,\cdots)$ assume a dispersion-relation form:
\begin{equation}
v^{\rm COV}_{ij}(q,\cdots)=\frac{1}{\pi}\int_{4m^2_{\pi}}^{\infty} dt' \;
\frac{g_{ij}(t',\cdots)}{\sqrt{t'}\;(t'+q^2)}\, ,
\label{eq:disper}
\end{equation}
and dots represent possible extra dependence on $\vec{p}\,'$ 
and $\vec{p}$ (kinetic energy in particular). 

The configuration-space PV TPE potential is obtained from the standard relation:
\begin{equation}
v(r)=\int \frac{d\vec{q}}{(2\pi)^3}\,e^{-i\vec{q}\cdot \vec{r}} \,v(q) \, .
\label{eq:fourier}
\end{equation}
Its expression thus reads:
\begin{eqnarray}
&&V(r,\vec{p}\,', \vec{p})=
i\,(\tau_1+\tau_2)^z\;\;(\vec{\sigma}_1 \times \vec{\sigma}_2)\!\cdot\!
[\vec{p},\; v_{44}(r,\cdots)]
\nonumber \\
&&\hspace*{3mm}
+\,(\tau_1+\tau_2)^z\;\;(\vec{\sigma}_1 - \vec{\sigma}_2)\!\cdot\!
\{\vec{p},\;  v_{34}(r,\cdots)\}
\nonumber \\
&&\hspace*{3mm}+i\,(\tau_1\times\tau_2)^z\;\;(\vec{\sigma}_1 + \vec{\sigma}_2)\!\cdot\!
 [\vec{p},\; v_{56}(r,\cdots)]
\nonumber \\
&&\hspace*{3mm}+\,(\tau_1-\tau_2)^z\;\;(\vec{\sigma}_1 + \vec{\sigma}_2)\!\cdot\!
\{\vec{p},\;  v_{75}(r,\cdots)\}
\nonumber \\
&&\hspace*{3mm}+2i\,(\tau_1 \!\times\! \tau_2)^z\;\;
\Big (\vec{\sigma}_1\!\cdot\! 
[ \vec{p},\, \vec{\sigma}_2\!\cdot\!\vec{l}\;
\frac{1}{r}\frac{d}{dr}v_{66}(r,\cdots) ] 
+ (\vec{\sigma}_1 \leftrightarrow   \vec{\sigma}_2)\Big)\;
\nonumber \\
&&\hspace*{3mm}-2\,(\tau_1\!-\!\tau_2)^z\;\;
\Big (\vec{\sigma}_1\!\cdot\! 
\{ \vec{p},\, \vec{\sigma}_2\!\cdot\!\vec{l}\;
\frac{1}{r}\frac{d}{dr}v_{85}(r,\cdots) \} 
+ (\vec{\sigma}_1 \leftrightarrow   \vec{\sigma}_2)\Big)\;
\,,
\label{eq:twopir}
\end{eqnarray}
where:
\begin{equation}
v^{\rm COV}_{ij}(r,\cdots)=\frac{1}{4\pi^2}\int_{4m^2_{\pi}}^{\infty} dt' \;
g_{ij}(t',\cdots)\;\frac{e^{-r\sqrt{t'}}}{r\;\sqrt{t'}}\,,
\end{equation}
while $\vec{p}\,'$ and $\vec{p}$, which dots account for, 
have now an operator character and should be placed respectively 
on the left and the right, in accordance with our conventions. 

The terms $V_{44}$, $V_{34}$, $V_{56}$ and $V_{75}$ appear at the first order 
in a $p/M$ expansion of the Lorentz invariants appearing in the expression of
the interaction. 
The two other ones, $V_{66}$ and $V_{85}$, appear at
the third order in $p/M$. They can therefore be considered as relativistic
corrections. Moreover, they only contribute when going beyond 
the  transitions between lowest partial wave states, i.e. $S$ to $P$, 
which generally dominate at low energy (where most PV data are available). 
For these two reasons,
the corresponding terms were discarded in the past, which we also do here. 
We however stress that these higher-order terms are necessary to get 
a full mapping of the $NN$ interaction, especially to discriminate 
transitions involving higher partial waves such as $^3P_1\!-\!\!\,^3D_1$ and $^3P_2\!-\!\!\,^3D_2$, 
$^3D_2\!-\!\!\,^3F_2$ and $^3D_3\!-\!\!\,^3F_3$, {\it etc}. It is also
noticed that the above  non-relativistic expansion only involves 
the nucleon external lines, as done in other approaches. None is 
made for internal nucleon lines where  big effects 
could arise. As our results presented in this paper only
retain part of the full relativistic structure, they will be denoted 
``covariant" to avoid  overstating this property.

We now consider the expressions of the $V_{44}$, $V_{34}$, 
$V_{56}$ and $V_{75}$ terms. It is first noticed  that $V_{44}$ and $V_{34}$ 
only receive contribution from the crossed diagram
while $V_{56}$ and $V_{75}$ also get some from the non-crossed one. 
In this case, one has therefore to worry about the removing 
of the iterated OPE and, especially, about the choice of the Green's function 
which enters the calculation. In Ref. \cite{desp_plb72},
the Green's function, $(2E_0-2E)^{-1}$, was used,  taking into account 
that the Schr\"odinger equation is linear in the energy of the system, $E_0$,
and assuming moreover that the kinetic energy of particles retains 
its relativistic form ($E=\sqrt{M^2+p^2}$ in the c.m.). 
In later works \cite{pirner_plb73,chem_npa74}, a Green's function 
more in agreement with a non-relativistic Schr\"odinger equation,
$E/(p_0^2-p^2)$, was instead used. The difference, 
a factor $2E/(E_0+E)$,   had minor numerical 
effects in the past calculations where cutoffs were introduced 
in the dispersion integrals, Eq. (\ref{eq:disper}). 
The difference however matters in unrestricted calculations. 
Dispersion-relation integrals diverge in the first case 
while they converge in the second one. With this last respect, it is noticed 
that the corresponding Green's function can be written in the following form, 
$E/(E_0^2-E^2)$, which rather evokes an equation 
with a quadratic-mass operator. Such an equation is known 
to provide solutions with a behavior in the relativistic domain 
better than the linear one \cite{amghar_npa03}. 
This is  the choice made here. 
Taking into account that the kinetic-energy dependence is small, 
and that we are interested in low-energy processes, 
the momentum in functions $g(q,\cdots)$ can be set to 0.  
In a similar way as Ref. \cite{desp_plb72} with details given in Appendix
\ref{app:gts}, the closed expressions for these functions are then obtained. 
Omitting dots that are no longer justified, they read:
\begin{eqnarray}
g_{44}(t') &=&\tilde{K} \frac{1}{2M} 
\Bigg (\frac{4q_{\pi}}{\chi^2}+\frac{H}{M^2} - 
G \Big (\frac{1}{M^2}+\frac{1}{\chi^2} \Big ) \Bigg)\, ,
\nonumber \\
g_{34}(t') &=& \tilde{K}\frac{1}{2M^3} 
\Bigg (G-\frac{H\;x^2}{x^2+4M^2q_{\pi}^2}\Bigg )\, ,
\nonumber \\
g_{56}(t')&=& -\tilde{K}\frac{1}{2M}\;\frac{H\;x}{x^2+4M^2q_{\pi}^2}
-\tilde{K}\frac{x}{M^2 \; m^2_{\pi}}\;
 {\rm arctg} \Big (\frac{m^2_{\pi}}{2Mq_{\pi}}\Big )
\nonumber \\
& &\!\! +\tilde{K}\int_{k_{-}^2}^{{k_{+}^2}} 
\frac{dk^2}{\sqrt{k^2t'\!-\!(m^2_{\pi}\!+\!k^2)^2}}\;
\frac {1}{2E\,(E\!+\!M)} 
\Bigg (\frac {x}{M^2}\!+\!\frac {k^2}{M\,(E\!+\!M)} \Bigg )\,,
\nonumber \\
g_{75}(t')&=&\tilde{K}\frac{4x^2}{M^2 \, m^2_{\pi}\,t'}\;
 {\rm arctg} \Big (\frac{m^2_{\pi}}{2Mq_{\pi}}\Big )+\tilde{K}\frac{2G}{Mt'}
\nonumber \\
& &\!\! -\tilde{K}\int_{k_{-}^2}^{{k_{+}^2}} 
\frac{dk^2}{\sqrt{k^2t'\!-\!(m^2_{\pi}\!+\!k^2)^2}}
\nonumber \\
& &\hspace*{3mm} \times  \frac{2}{E(E\!+\!M)}\Bigg (
 \frac{x^2}{M^2t'}\!+\!\frac{2E\,x}{Mt'}
\!-\!\frac{k^2}{M(E\!+\!M)}
\!-\! \frac{k^4}{M(E\!+\!M)t'}  \Bigg )\,,
\label{eq:gcad}
\end{eqnarray}
where:
\begin{eqnarray}
\tilde{K}&=&\frac{g^3_{\pi\,NN}\; h^1_{\pi}}{32\pi\sqrt{2}}\,,
\nonumber \\
q_{\pi}&=&\sqrt{\frac{t'}{4}-m^2_{\pi}},\hspace*{1cm} \chi^2=M^2-\frac{t'}{4}, 
\hspace*{1cm} x=\frac{t'}{2}-m^2_{\pi}\, ,
\nonumber \\
H&=&2\sqrt{\frac{x^2+4M^2q_{\pi}^2}{t'}} \;\;{\rm ln} 
\Bigg (\frac{\sqrt{x^2+4M^2q_{\pi}^2} + q_{\pi}\sqrt{t'}   }{
      \sqrt{x^2+4M^2q_{\pi}^2} - q_{\pi}\sqrt{t'}}
\Bigg )\, ,
\nonumber \\
G&=&\frac{2x}{\chi}\;{\rm arctg} \Big (\frac{2q_{\pi}\chi}{x}\Big)\;\;\;
[\chi^2\geq0],\;\;\;
=\frac{x}{\sqrt{\!-\!\chi^2}}\;{\rm ln}
\Bigg (\frac{x+2q_{\pi}\sqrt{\!-\!\chi^2}}{x-2q_{\pi}\sqrt{\!-\!\chi^2}}\Bigg
)\;\;\;[\chi^2\leq0]\, ,
\nonumber \\
k_{\pm}^2 &=& x\pm q_{\pi}\sqrt{t'},\hspace*{1cm} E=\sqrt{M^2+k^2}\, .
\label{eq:defs}
\end{eqnarray}

Looking at the asymptotic behavior of the functions $g_{ij}(t')$ 
for large $t'$, it is noticed that the dominant contributions 
of individual terms, $\propto  t'^{1/2}$, as well as the constant ones, 
cancel (see Appendix \ref{app:green} for some detail). One is thus left 
with the following contributions:
\begin{eqnarray}
g_{44}(t')_{t' \rightarrow \infty}&=&\tilde{K} \frac{2}{M \sqrt{t'}}
\Bigg ( {\rm ln}  \Big (\frac{t'}{M^2} \Big ) -1\Bigg )\,,
\nonumber \\
g_{34}(t')_{t' \rightarrow \infty}&=& \tilde{K} \frac{2}{M \sqrt{t'}}
\Bigg ( {\rm ln}  \Big (\frac{t'}{M^2} \Big ) -1\Bigg )\,,
\nonumber \\
g_{56}(t')_{t' \rightarrow \infty}&=& - \tilde{K}\frac{1}{M \sqrt{t'}}
\Bigg (\frac{5}{8 }\,{\rm ln}  \Big (\frac{t'}{M^2} \Big ) 
+\frac{15}{8}-\frac{3}{2}\,{\rm ln}(2)\Bigg )\,,
\nonumber \\
g_{75}(t')_{t' \rightarrow \infty}&=& \tilde{K}\frac{1}{M \sqrt{t'}}
\Bigg ( \frac{9}{4}\, {\rm ln}  \Big (\frac{t'}{M^2} \Big ) 
-\frac{1}{4}+\,{\rm ln}(2)\Bigg )\, .
\label{eq:larget}
\end{eqnarray}
In all cases, the functions $g_{ij}(t')$ behave asymptotically 
as $t'^{-1/2}$, up to log terms, ensuring the convergence 
of the integrals given in Eq. (\ref{eq:disper}). This result is important 
as it allows one to consider the dispersion approach as a benchmark, thus 
providing information about contributions that are ascribed to LEC's 
as well as possible higher-order corrections in other approaches.


\section{Two-pion exchange from the EFT approach}
\label{EFT}

\subsection{EFT approach}
\begin{figure}[htb]
\begin{center}
\mbox{ \epsfig{ file=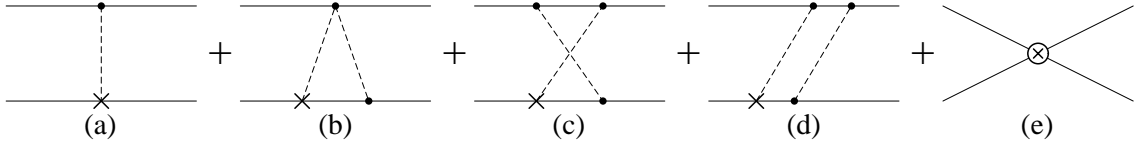, width=15cm}}
\end{center}
\caption{One and two-pion exchange in the EFT approach: 
The contributions from the OPE (a), from the triangle TPE (b), 
from the crossed TPE (c), from the non-crossed TPE (d) and 
from the contact term (e). See the caption of Fig. \ref{fig:covar} 
for further comments.
\label{fig:eft}}
\end{figure} 
The PV TPE at NNLO has been calculated in the EFT approach 
by Zhu {\it et al.} \cite{zhu_npa05}.
It contains two components which, in their notations, 
involve functions $\tilde{C}^{2\pi}_2(q)$ and $C^{2\pi}_6(q)$ and correspond here 
to the components $V_{44}$ and $V_{56}$ of the more complete interaction 
given by Eq. (\ref{eq:twopi}). The contributions being accounted for 
correspond to the diagrams shown in Fig. \ref{fig:eft}. 
Their expressions for the finite-range part and the associated contact term 
have been obtained in the maximal-subtraction (MX) scheme.  
Factorizing out the spin-isospin dependence and taking into account 
corrections made since then \cite{hyun_plb07}, 
they read:~\footnote{We have followed our own conventions, 
as the conventions used to present the final results in Ref. \cite{zhu_npa05} 
differ from that ones defined by the same authors at an earlier stage.}
\begin{eqnarray}
v_{44}^{\rm EFT}(q)
&\!=\!& -4\sqrt{2}\;\pi \frac{h^1_\pi}{\Lambda^3_\chi}
 \Big (g_A^3\,L(q)\Big )\, , 
\nonumber \\
v_{56}^{\rm EFT}(q)
&\!=\!&-\sqrt{2}\;\pi \frac{h^1_\pi}{\Lambda^3_\chi}
\Bigg ( g_A\;L(q) 
-g_A^3\;\Big(3L(q)\!-\!H(q)\Big)\Bigg )
\, , \label{eq:eft}
\end{eqnarray}
where the scale $\Lambda_\chi$ is roughly given by 
$\Lambda_\chi=4\pi f_{\pi} \simeq 4\pi \,g_A\,M/g_{\pi NN}\simeq 1$ GeV.
The $L(q)$ and $H(q)$ functions are defined as:
\begin{eqnarray}
&&L(q) =\frac{\sqrt{q^2 \!+\! 4 m_\pi^2}}{q}\,
\ln \!\left( \frac{\sqrt{q^2\!+\!4 m_\pi^2} + q}{2 m_\pi} \right)
=\frac{\sqrt{q^2 \!+\! 4 m_\pi^2}}{2\,q}\,
\ln \!\left( \frac{\sqrt{q^2\!+\!4 m_\pi^2} + q}{\sqrt{q^2\!+\!4 m_\pi^2} -
q}\right)\,,
\nonumber \\
&&H(q) = \frac{4m_\pi^2}{q^2\!+\!4m_\pi^2}\;L(q)\, .
\label{eq:lq} 
\end{eqnarray}
The detail of the contributions corresponding to the diagrams (b), (c) and (d)
in Fig. \ref{fig:eft} can be found in Appendix \ref{app:eft}. The above
terms entering the interaction should be completed by contact terms:
\begin{eqnarray}
&&v^{CT}_{44}= C_{44}\, ,
\nonumber \\
&& v^{CT}_{56}= C_{56}\, .
\end{eqnarray}
The contributions from two-pion exchange to these contact terms are also given  
in Appendix \ref{app:eft}, where it is seen that they require some
renormalization. The sum of the EFT two-pion-exchange and contact contributions 
should be well determined, but  how it is split between the two terms is not.  
In the minimal-subtraction ($\overline{ \mbox{\rm MS}}$) 
scheme,\footnote{What we denoted by MN the scheme in a previous work 
\cite{hyun_plb07} actually corresponds to the $\overline{ \mbox{\rm MS}}$ scheme, 
hence the change of name adopted here.} 
the part of the contact term which is
proportional to $1\!+\!{\rm ln} (\mu/m_{\pi})$ (see Appendix \ref{app:eft}) 
is shifted to the EFT two-pion-exchange part. In such a case, 
the term ``1" cancels the function $L(q)$ at $q=0$ and the log term, 
for $\mu \geq m_{\pi}$, changes the overall sign of the potential 
in the low-$q$ range.

\subsection{Relation to the time-ordered-diagram approach}
\begin{figure}[htb]
\begin{center}
\mbox{ \epsfig{ file=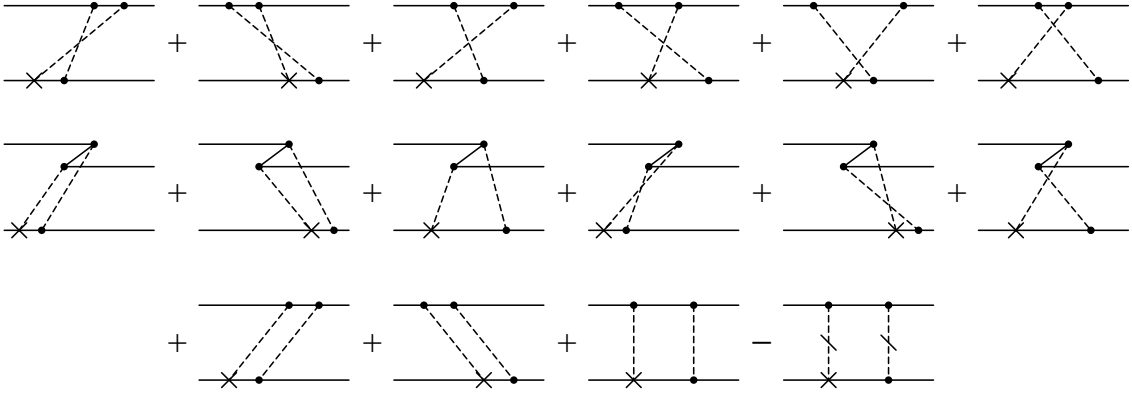, width=15cm}}
\end{center}
\caption{Two-pion exchange in the time-ordered diagram approach: 
The contributions from the crossed diagrams (the first line),
from the Z-type contributions (the second line, identified 
as triangle diagrams in a different approach), from the non-crossed diagrams 
(the diagrams on the third line). 
The last two diagrams differ in that the first of them involves
a meson propagator with off-energy shell contributions, 
which are omitted in the other one (what is reminded by a back slash 
on the meson lines). This last contribution, which arises from the iterated OPE, 
has to be subtracted from the previous one. The discrepancy 
involves a factor, $E_0-k^2/M$, which cancels a similar factor entering 
the denominator of the Green function 
($\vec{k}$ can be identified as the loop momentum). 
See the caption of Fig. \ref{fig:covar} for further comments.
\label{fig:timeord}}
\end{figure} 
When integrating the EFT expressions of the TPE interaction 
over the time component of the integration variable entering some loop, 
it is expected that one should recover expressions obtained from considering
time-ordered (TO) diagrams in the non-relativistic limit ($\vec{v}\rightarrow 0$ 
where $\vec{v}$ is the nucleon velocity $\vec{v}\simeq \vec{p}/M$).
As this check was quite useful in determining the correct expressions given in
the previous subsection, we give below the raw expressions obtained 
from the contribution of these time-ordered diagrams, which are shown in 
Fig. \ref{fig:timeord}. 
Starting from the elementary $\pi NN$ interaction,  
$g_{\pi NN}\,\vec{\sigma}\cdot \vec{k}/(2M\,\sqrt{\omega_k})$ 
(with $\omega_k=\sqrt{m_\pi^2+k^2} $), one gets:
\begin{eqnarray}
v^{\rm TO}_{44}(q)&\!=\!&\frac{g_{\pi NN}^3\; h^1_{\pi}}{4\sqrt{2}\,M^3}\;\;
\nonumber \\
&& \times \int \frac{d\vec{k}}{(2\pi)^3}
\frac{k^2\!-\!(\vec{k}\cdot\hat{q})^2}{\omega_i\;\omega_j }\;
\Bigg (\frac{1}{\omega_i^2\,(\omega_i\!+\!\omega_j)}
+\frac{1}{(\omega_i\!+\!\omega_j)\,\omega^2_j}  
+\frac{1}{\omega_i\,(\omega_i\!+\!\omega_j)\,\omega_j} \Bigg)\,,
\nonumber \\
v^{\rm TO}_{56}(q)&\!=\!&\frac{g_{\pi NN}^3\; h^1_{\pi}}{4\sqrt{2}\,M^3}\;\;
\Bigg( 
\frac{1}{2}\int \frac{d\vec{k}}{(2\pi)^3}
\frac{1}{\omega_i\,\omega_j\,(\omega_i\!+\!\omega_j) }
\nonumber \\
&& -\frac{1}{2} \int \frac{d\vec{k}}{(2\pi)^3}\;
\frac{k^2\!-\!\frac{q^2}{4}}{\omega_i\;\omega_j }\;
\Bigg (\frac{1}{\omega_i^2\,(\omega_i\!+\!\omega_j)}
+\frac{1}{(\omega_i\!+\!\omega_j)\,\omega^2_j}  
+\frac{1}{\omega_i\,(\omega_i\!+\!\omega_j)\,\omega_j} \Bigg)\Bigg)\,,
\label{eq:vtord}
\end{eqnarray}
where $\omega_i=\sqrt{m^2_{\pi}+(\vec{k}+\vec{q}/2)^2}$, 
$\omega_j=\sqrt{m^2_{\pi}+(\vec{k}-\vec{q}/2)^2}$. 
The three integrals involve the contributions successively 
from the crossed diagrams (the first line of Fig. \ref{fig:timeord}),
Z-type ones (the second line of Fig. \ref{fig:timeord}) 
and both crossed and non-crossed diagrams
(the first and third lines of Fig. \ref{fig:timeord}). The Z-type diagrams 
are calculated assuming a pseudo-scalar coupling, consistently 
with the dispersion-relation approach used independently, 
and retaining the lowest non-zero term in a $1/M$ expansion. 
As is known, the contribution alone violates chiral symmetry 
(see, for instance, Ref. \cite{coon_prc86}). The expected symmetry 
is restored by further contributions, which can be  calculated 
in the same formalism (see some detail in Sec. \ref{relation_dif}).
Contributions of all diagrams in Fig. \ref{fig:timeord} diverge 
but they contain a well-defined part that can be analytically calculated. 
Interestingly, the expressions 
so obtained can be cast into the form of dispersion integrals. 
This property stems from considering a complete set of  
topologically-equivalent time-ordered diagrams. This writing 
is interesting as it greatly facilitates the comparison 
with the expressions obtained from a covariant approach, 
which evidences the same form. It is thus found 
that the different integrals in Eqs. (\ref{eq:vtord}) read:
\begin{eqnarray}
&&\int d\vec{k}\;
\frac{k^2\!-\!(\vec{k}\cdot\hat{q})^2}{\omega_i\;\omega_j }\;
\Bigg (\frac{1}{\omega_i^2\,(\omega_i\!+\!\omega_j)}
+\frac{1}{(\omega_i\!+\!\omega_j)\,\omega^2_j}  
+\frac{1}{\omega_i\,(\omega_i\!+\!\omega_j)\,\omega_j} \Bigg)
\nonumber \\
&&\hspace*{1cm}=4\pi\Big(1-L(q)\Big)+\int d\vec{k}\,\frac{k^2}{\omega_k^5}
=\pi\int_{4m_\pi^2}^{\infty} dt' \;
\frac{2\sqrt{t'-4\,m_\pi^2}}{\sqrt{t'}\;(t'+q^2)}\,,
\nonumber \\
&&\int d\vec{k}\;
\frac{1}{\omega_i\,\omega_j\,(\omega_i\!+\!\omega_j) }
\nonumber \\
&&\hspace*{1cm}=2\pi\Big(1-L(q)\Big)+\frac{1}{2}\int d\vec{k}\,\frac{1}{\omega_k^3}
=\pi\int_{4m_\pi^2}^{\infty}  dt' \;
\frac{\sqrt{t'-4\,m_\pi^2}}{\sqrt{t'}\;(t'+q^2)}\,,
\nonumber \\
&& \int d\vec{k}\;
\frac{k^2\!-\!\frac{q^2}{4}}{\omega_i\;\omega_j }\;
\Bigg (\frac{1}{\omega_i^2\,(\omega_i\!+\!\omega_j)}
+\frac{1}{(\omega_i\!+\!\omega_j)\,\omega^2_j}  
+\frac{1}{\omega_i\,(\omega_i\!+\!\omega_j)\,\omega_j} \Bigg)\Bigg)
\nonumber \\
&&\hspace*{1cm}=
2\pi\Bigg(3\Big(1-L(q)\Big)+H(q)\Bigg)
+\frac{3}{2}\int d\vec{k}\,\frac{k^2}{\omega_k^5}
\nonumber \\
&&\hspace*{1cm}=\pi\int_{4m_\pi^2}^{\infty} dt'\;\;
\frac{3(t'\!-\!4m_\pi^2)+4m_\pi^2}{\sqrt{t'}\;\sqrt{t'\!-\!4m_\pi^2}\;(t'\!+\!q^2)}\, .
\label{eq:ordered}
\end{eqnarray}
Due to the divergent character of these integrals, the above equalities hold up 
to some constant. This does not however affect the $q^2$-dependent part. 
One can thus remove an infinite contribution so that the interaction 
takes a definite value at some $q^2$. A particular interesting choice 
is to subtract a part so that the remaining one, 
which contains the most physically relevant part, cancels at $q^2\!=\!0$. 
By looking at this quantity, one can usefully compare different approaches. 
To some extent the slope with respect to $q^2$ at $q^2\!=\!0$ 
provides information on the sign  and  the strength of the interaction 
at finite distances. The $L(q)$ function in the above equalities thus points 
to a configuration-space interaction with an opposite sign at finite distances.
The divergent part is not without interest however. It tells us in which
direction the (short-range) subtracted interaction is likely to contribute. 
By integrating out the contribution $t'\geq \tilde{\Lambda}^2$ at the r.h.s. 
of Eqs. (\ref{eq:ordered}) in the small $q$ limit, one successively gets 
the approximate factors $4\pi\, {\rm ln}(\tilde{\Lambda}/2m_{\pi})$, 
 $2\pi\, {\rm ln}(\tilde{\Lambda}/2m_{\pi})$, 
and $6\pi\,{\rm ln}(\tilde{\Lambda}/2m_{\pi})$. This suggests two observations. 
On the one hand,  for large enough $\tilde{\Lambda}$, the r.h.s. has a sign opposite 
to that one given by the $L(q)$ term at the l.h.s., 
confirming the observation in configuration space. 
On the other hand, the above factor allows one to make some relation 
with the EFT interaction calculated in the minimal-subtraction scheme,
which involves similar log terms (with $\tilde{\Lambda}$ replaced by  $\mu$). 

Comparing the above expressions,  Eqs. (\ref{eq:ordered}), 
with the previous EFT ones, Eqs. (\ref{eq:eft}),
it is found that the $q^2$ dependences are very similar. 
Assuming the choice  $\Lambda_\chi=4\pi \,g_A\,M/g_{\pi NN}$, 
an identity is actually found for the potential, $v_{44}(q)$,
as well as the second term for the other potential, $v_{56}(q)$. 
For the first term in $v_{56}(q)$, which can be associated 
with a triangle-type diagram, the time-ordered diagram approach 
used here gives a factor $g_A^3$ instead of  $g_A$ 
as directly obtained from the Weinberg-Tomozawa term. 
This discrepancy points to the fact that the approach misses 
some contribution. As already mentioned, this will be discussed 
in more details in Sec. \ref{relation_dif} when making a comparison 
with the expressions obtained from the covariant formalism.

\section{Relation of the EFT approach to the covariant one}
\label{sec_relation}
We here discuss similarities and differences between the expressions 
obtained from the full covariant approach and the EFT one presented 
in sections \ref{COV} and \ref{EFT} respectively.
\subsection{Similarities (large-$M$ limit)} 
\label{relation_sim}
To make a comparison of the EFT and time-ordered-diagram approaches 
with the covariant one, the first step is to derive expressions 
in the large-$M$ limit for the last case. Taking this limit 
in the simplest-minded way for the $H$ and $G$ functions 
given in Eq. (\ref{eq:defs}), one gets:
\begin{eqnarray}
H_{M\rightarrow \infty}&=&4\,q_{\pi}= 2\sqrt{t'-4m^2_{\pi}} \, ,
\nonumber \\
G_{M\rightarrow \infty}&=& \frac{(t'-2m^2_{\pi})}{M}
\Bigg(\frac{\pi}{2}-\frac{x}{2q_{\pi}M}\Bigg) \simeq \pi \frac{x}{M}\,.
\label{eq:GHlimit}
\end{eqnarray}
Inserting the above limits in Eqs. (\ref{eq:gcad}), one finds:
\begin{eqnarray}
g_{44}(t')_{M\rightarrow \infty} &\!=\!& \tilde{K}\;\frac{4q_{\pi}}{M^3} 
=\tilde{K}\; \frac{2\sqrt{t'\!-\!4m^2_{\pi}}}{M^3} \, , 
\nonumber \\
g_{34}(t')_{M\rightarrow \infty} &\!=\!& \tilde{K}\;\frac{\pi}{M^4} \,
\frac{x}{2}=\tilde{K}\;\frac{\pi}{M^4}\frac{t'\!-\!2m^2_{\pi}}{4} \, , 
\nonumber \\
g_{56}(t')_{M\rightarrow \infty}
&\!=\!&-\tilde{K}\;\frac{x}{q_{\pi}\,M^3}  
=-\tilde{K}\; \frac{(t'\!-\!2m^2_{\pi})}{M^3\,\sqrt{t'\!-\!4m^2_{\pi}}} \, , 
\nonumber \\
g_{75}(t')_{M\rightarrow \infty}
&\!=\!&\tilde{K}\;\frac{\pi}{M^4} \;
(\frac{t'-4m^2_{\pi}}{16}+\frac{3}{2}\, \frac{t'-2m^2_{\pi}}{4})\,.
\label{eq:glargeM}
\end{eqnarray}
In obtaining the above expressions, one has taken into account 
that the integral in Eq. (\ref{eq:gcad}) for $g_{56}(t')$ 
has a higher  $1/M$ order ($1/M^4$). 
The case of $g_{75}(t')$ is more complicated as individual
contributions of $1/M^2$ and $1/M^3$ order cancel. 
In any case, we notice that the large-$M$ limit does not commute 
with the large-$t'$ limit (compare with the results given 
in Eq. (\ref{eq:larget})). 
Differences from the  ``covariant" results are therefore expected when considering
short distances, where the last limit is relevant.  
The insertion of the above limits 
in the dispersion-relation integrals, Eqs. (\ref{eq:disper}), 
allows one to recover the expressions of the time-ordered-diagram approach 
at the lowest order, $1/M^3$, for interactions $V_{44}$ and $V_{56}$: 
\begin{eqnarray}
&&v_{44}(q)_{M \rightarrow \infty}=
\frac{g_{\pi NN}^3\; h^1_{\pi}}{16\sqrt{2}\;\pi^2\,M^3}\;\;
\int_{4m^2_{\pi}}^{\infty} dt' \;
\frac{\sqrt{t'-4\,m^2_{\pi}}}{\sqrt{t'}\,(t'+q^2)}\, ,
\nonumber \\ 
&&v_{56}(q)_{M\rightarrow \infty}=
-\;\frac{g_{\pi NN}^3\; h^1_{\pi}}{32\sqrt{2}\;\pi^2\,M^3}\;
\int_{4m^2_{\pi}}^{\infty} dt' \;
\frac{(t'\!-\!2m^2_{\pi})}{\sqrt{t'}\;\sqrt{t'\!-\!4m^2_{\pi}}\;(t'\!+\!q^2)}\,.
\label{eq:vlargeM0}
\end{eqnarray}
For the interaction $V_{56}$, it is noticed that the two terms 
involving integrands proportional to $-(t'\!-\!4m^2_{\pi})$ 
and $3(t'\!-\!4m^2_{\pi})+4m^2_{\pi}$ in Eq. (\ref{eq:vtord}) 
(together with Eq. (\ref{eq:ordered})) combine to give 
the factor proportional to $(t'\!-\!2m^2_{\pi})$ 
appearing in the large-$M$-limit expression of $g_{56}(t')$, 
Eq. (\ref{eq:glargeM}). 

Expressions for both $V_{44}$ and $V_{56}$ 
can be cast into a form that facilitates the comparison
with Zhu {\it et al.}'s work \cite{zhu_npa05}. In this order, 
a factor  $g_{\pi NN}^3 /(4\,\pi g_A M)^3 $, which can be identified 
as the $1/\Lambda^3_{\chi}$ factor in their work, is partly 
factored out. Infinities present in the dispersion integrals 
are temporarily ascribed to LEC's, knowing that these ones 
should be finite in practice. We thus have:
\begin{eqnarray}
&&v_{44}(q)_{M \rightarrow \infty}= v^{\rm LM}_{44}(q) + C'_{44} \, ,
\nonumber \\ 
&&v_{56}(q)_{M\rightarrow \infty}=v^{\rm LM}_{56}(q) +C'_{56} \, ,
\nonumber 
\end{eqnarray}
with:
\begin{eqnarray}
&& v^{\rm LM}_{44}(q)=
-4 \sqrt{2}\,\pi \frac {g_{\pi NN}^3\;\;h^1_{\pi}}{(4\,\pi g_A M)^3} \;
\; g_A^3\;L(q) \, , 
\nonumber \\ 
&& v^{\rm LM}_{56}(q)=
- \sqrt{2}\,\pi\frac {g_{\pi NN}^3 \;h^1_{\pi}}{(4\,\pi g_A M)^3} \;
\Bigg ( g_A^3\;L(q) 
-g_A^3\;\Big(3L(q)\!-\!H(q)\Big)
\Bigg )\, .
\label{eq:vlargeM}
\end{eqnarray}
The only significant discrepancy with  Zhu {\it et al.}'s work, 
Eqs.  (\ref{eq:eft}),
concerns the first term of $V_{56}$, which contains 
a factor $g_A^3$ instead of $g_A$, confirming the observation 
already made in the time-ordered-diagram approach. 

\subsection{Differences}
\label{relation_dif}
After having shown how the expressions of the EFT (or the time-ordered-diagram) 
approach can be obtained from the covariant ones for the gross features, 
we now examine the differences.

The first difference concerns the convergence properties 
of the integral expressions for the potentials $v_{ij}(q)$, 
Eqs. (\ref{eq:disper}). 
While the EFT ones do not converge (infinite LEC's), 
those components produced by the crossed-box diagram, $V_{44}$ and $V_{34}$ 
in the covariant approach always converge. 
This also holds for the crossed-box part of the other components 
$V_{56}$ and $V_{75}$ but, in these cases, one has to consider a further 
contribution from the non-crossed box diagram.   
The convergence crucially depends on the way the iterated OPE 
is calculated but there is one choice, quite natural actually, 
which provides convergence as good as for the crossed-box diagram. 
Though it does not really make sense physically 
to integrate dispersion integrals over $t'$ up to $\infty$, 
expressions so obtained provide a reliable benchmark, 
as far as the same physics is implied. 

The second difference concerns the number of components. 
The covariant approach involves many more than the EFT one at NNLO 
(6 instead of 2).
The extra ones imply some recoil effect and have a non-local character. 
They are of higher order in a $1/M$ expansion but, instead, 
in the large-$t'$ limit, they compare to the other ones, 
see Eqs. (\ref{eq:larget}).

The third difference has to do with the large-$M$ limit 
of the covariant approach, Eqs. (\ref{eq:gcad}), which allows one 
to recover the structure of the EFT results. 
The way this limit is taken in the $H$ or $G$ functions, 
or in the factor multiplying the first quantity, is quite rough. 
It assumes approximations like  $x^2+4M^2q_{\pi}^2 \simeq 4M^2q_{\pi}^2 $ 
($2q_{\pi}M \geq x$).
Actually, due to the small but finite value of the pion mass, 
there is a very little range of $t'$ values 
($t'\!-\!4m^2_{\pi} \leq m^4_{\pi}/M^2$) where this approximation is not valid. 
The correction, which disappears in the chiral limit 
(zero pion mass), could affect the long-range part of the interaction.

The fourth difference involves chiral symmetry and related properties. 
Contrary to what is sometimes thought, fulfilling these properties 
in calculating the TPE contribution in the covariant approach is possible. 
This however supposes some elaboration, requiring that a description 
of the $N\bar{N}\rightarrow \pi\pi$ transition amplitude entering 
the dispersion relations is consistent 
with chiral symmetry. In the instance of the Paris model 
for the $NN$ strong interaction \cite{paris}, this amplitude 
could be related to experimental data. For the $NN$ weak interaction, 
the strong amplitude was instead modeled from the contribution 
of a few nucleon resonances in the $s$-channel \cite{chem_npa74}. 
This can be essentially achieved by adding to the nucleon intermediate state
retained here the contribution of the $\Delta(1232\,{\rm MeV})$ resonance.
This one, in the dispersion-relation formalism, suppresses 
the low-energy $N\pi \leftrightarrow N\pi$ strong-transition amplitude, 
otherwise dominated by a well-known large Z-type contribution 
inconsistent with chiral symmetry. This part, which involves two pions 
in an isosinglet state (with the $\sigma$-meson quantum numbers), 
is irrelevant here however. Its contribution is suppressed, 
in accordance with the Barton theorem \cite{barton_nc61} 
which states that  the exchange of scalar and pseudo-scalar neutral mesons 
does not contribute to the PV $NN$ interaction (assuming $CP$ conservation). 
This feature largely explains why the PV TPEP contribution 
has not been found as important as originally expected, on the basis 
of the strong-interaction case \cite{chem_npa74}.
There is another part  that is of interest here. It decreases 
the Z-type contribution (the first term of $V_{56}(q)$ in Eq. (\ref{eq:vtord}))
by an amount which corresponds to changing the factor $g_A^3$ 
into $g_A$  in the first term of $V_{56}(q)$ in Eq. (\ref{eq:vlargeM}). 
This can be checked in the simplest non-relativistic quark model 
according to the relation: 
\begin{equation}
g^2_{\pi NN}-
\frac{2g^2_{\pi N\Delta}}{9}=(1-\frac{16}{25})\,g^2_{\pi NN}
=\frac{9}{25}\,g^2_{\pi NN}=\frac {g^2_{\pi NN}}{g_A^2}\,,
\end{equation}
which shows that the contribution of nucleon resonances to the $\pi N $
scattering amplitude can be accounted for by dividing the intermediate 
nucleon contribution by a factor $g_A^2$. A somewhat different 
but better argument is based on the Adler-Weisberger sum rule 
\cite{adler-weis}. This one, which does not involve any non-relativistic 
limit, can be cast into the form $1-\int \cdots=1/g^2_A$, where 
the integral involves the off-mass-shell pion-proton  total cross sections.
For simplicity, we did not retain here the above $\Delta$ contribution, 
possibly improved for other resonances. 
We nevertheless keep in mind from the previous considerations 
that the discrepancy between the EFT and the covariant approaches, 
which was noticed for the contribution of the triangle diagram 
in the former one (a factor $g_A$ instead of $g_A^3$), 
could be removed by completing the latter one. 
The corresponding contribution, considered in Ref. \cite{chem_npa74},
amounts to (10$-$20)\% of the one retained here.

A last remark concerns the comparison of the TPE with the $\rho$-meson 
exchange. For a part, the first one was discarded in the past 
due to possible double counting with the second one. 
With this respect, we notice that the ratio of the $V_{44}$ and $V_{34}$
components, which could contribute to PV effects in $pp$ scattering
($^1S_0\!-\!\!\,^3P_0$ transition amplitude), is very much like the ratio 
of the local and non-local parts of a standard $\rho$-exchange contribution. 
There is some relationship between this result and the fact that 
the pion cloud produces a contribution to the anomalous magnetic moment 
of the nucleon, which compares to the physical one. 
The problem is different for the other PV transition, 
$^3S_1\!-\!\!\,^3P_1$, where the $V_{56}$ component could contribute. 
The corresponding charged-$\rho$ exchange of interest in this case 
is governed by the PV coupling $h'^1_{\rho}$, which was predicted 
to vanish in the DDH work \cite{DDH_ap80}. A small value ($-0.7\times 10^{-7}$) 
was obtained by Holstein \cite{holstein_pr23} by considering 
a pole model. A larger value ($-$(2$-$3)$\times 10^{-7}$) was obtained later on 
by Kaiser and Meissner \cite{km_npa499}  using a soliton model. 
In any case, these values lead to negligible effects. We however observe 
that the PV $\pi NN$ coupling constant $h^1_{\pi}$ is also small 
in the same models, most of its possible larger predicted value 
being due to the contribution of strange quarks \cite{desp_phrep98}. 
Sizable values of $h'^1_{\rho}$ could thus be expected. On the basis 
of a dynamical model considering the $\rho$ meson as a two-pion resonance, 
one cannot exclude values of  $h'^1_{\rho}$ in the range of 
(5$-$10)$\sqrt{2}\,h^1_{\pi}$, a relation that the results of
Kaiser and Meissner roughly verify. Contrary to the previous values, 
the last ones could lead to some double counting and it is likely 
that the $h'^1_{\rho}$ contribution to PV effects is then largely accounted for 
by the two-pion exchange considered in this work. It is conceivable that a
similar conclusion holds for the other isovector coupling, $h^1_{\rho}$, 
which contributes to the PV effects in $pp$ scattering discussed above.
Another aspect of the comparison with a $\rho$ exchange 
concerns the range of the TPE, which was presented in Ref. \cite{zhu_npa05} 
as a medium one. This could apply to the exchange 
of two pions in a $S$ wave ($\sigma$ meson), which contributes 
to the strong $NN$ interaction but is absent in the weak case as already
mentioned. The TPE of interest here involves two pions in a $P$ wave 
with the quantum numbers of the $\rho$ meson. Due to a centrifugal barrier 
factor, the TPE contribution is shifted to values of $t'$ higher 
than for a $S$ wave, making the range of its contribution closer 
to a $\rho$-exchange one. Some numerical illustration is given 
in next section.

\section{Numerical comparison of potentials}
\begin{figure}[htb]
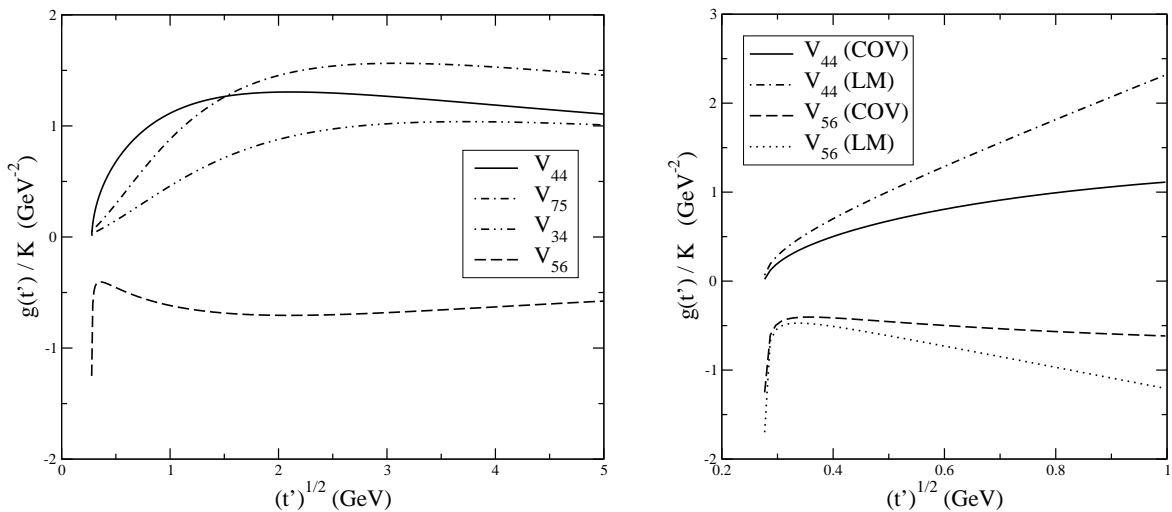

\begin{center}
\mbox{ \epsfig{ file=gtsl5.eps, width=8cm} \hspace*{0.5cm}
\epsfig{ file=gtgml1.eps, width=6.7cm}}
\end{center}
\caption{Spectral functions $g(t')$ in units of GeV$^{-2}$, 
with the coefficient $\tilde{K}$ factored out, and represented 
as a function of $t'^{1/2}$ to better emphasize the low-$t'$ range. 
Left panel: for $t'^{1/2}$ from threshold to  $5\,{\rm GeV}$ 
for all functions, to show the relative importance of the various components 
and the beginning of the onset of the asymptotic behavior, $t'^{-1/2}$. 
Right panel: from threshold to  $1\,{\rm GeV}$ for functions 
$g_{44}(t')$ and $g_{56}(t')$ together with their large-$M$ limits, 
to check the validity of this approximation.  
\label{fig:gts}}
\end{figure} 
We consider in this section various numerical aspects of the potentials
presented in the previous one. They successively concern the spectral
functions, $g_{ij}(t')$, the potentials in momentum space, 
$v_{ij}(q)$, and the potentials in configuration space, $v_{ij}(r)$. 
In most cases, we directly compare the results of the covariant approach (COV)
with those obtained from it in  the large-$M$ limit (LM), 
Eqs.  (\ref{eq:vlargeM}). This comparison is more meaningful
than the one with the EFT potential (EFT), Eqs.  (\ref{eq:eft}), 
as it is not biased by the choice of the factor $\Lambda_{\chi}$ 
and by the difference of a factor $g_A$, instead of $g_A^3$, 
in part of the contribution to $V_{56}$, of which origin is understood 
in any case.
Before entering into details, we notice that the various components 
of the interaction have a local character for some of them 
($V_{44}$ and $V_{56}$) and a non-local one for the other ones 
($V_{34}$ and $V_{75}$). 
At low energy, however, it turns out that one of the contributions 
involving the factor $\vec{p}$ or $\vec{p}\,'$ in their expression,
Eq. (\ref{eq:twopir}), is small. Moreover, with our conventions, 
the spin-isospin factors give the same values for the $z$ component.  
The various components can then be usefully compared, independently 
of their local or non-local character, that is what we do here. 
Numerical results assume the following values:
$g_{\pi NN}=13.45, \; g_A=1.2695,\; M= 938.919 {\rm MeV},\; 
m_{\pi}=138.039  {\rm MeV}, \; m_{\rho}=771.1 {\rm MeV}$.
\begin{figure}[htb]
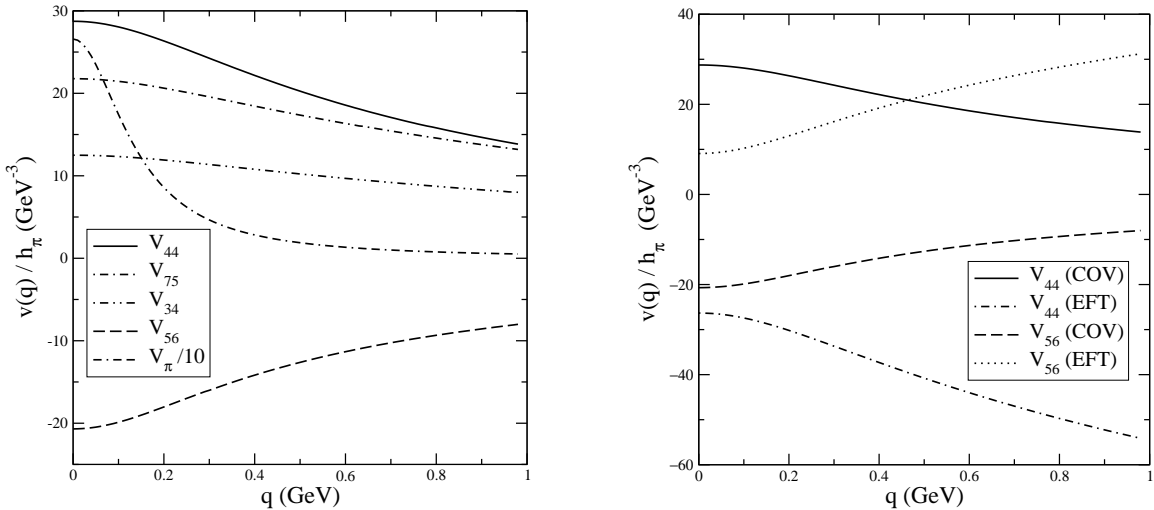

\begin{center}
\mbox{ \epsfig{ file=potqfus1.eps, width=7cm} \hspace*{1cm}
\epsfig{ file=potqfugms1.eps, width=7cm}}
\end{center}
\caption{Potentials $v_{44}(q)$, $v_{34}(q)$, $v_{56}(q)$ and $v_{75}(q)$ 
(together with the OPE one divided by 10) for $q$ ranging from 0 to 1 GeV 
(left panel) and their EFT counterpart for $v_{44}(q)$ and  $v_{56}(q)$ 
(right panel). Ingredients entering the EFT results are specified in the text.
Notice that the EFT and  ``covariant" versions 
of a given component of the potential have opposite signs.
\label{fig:vqs}}
\end{figure} 
\begin{figure}[htb]
\begin{center}
\mbox{ \epsfig{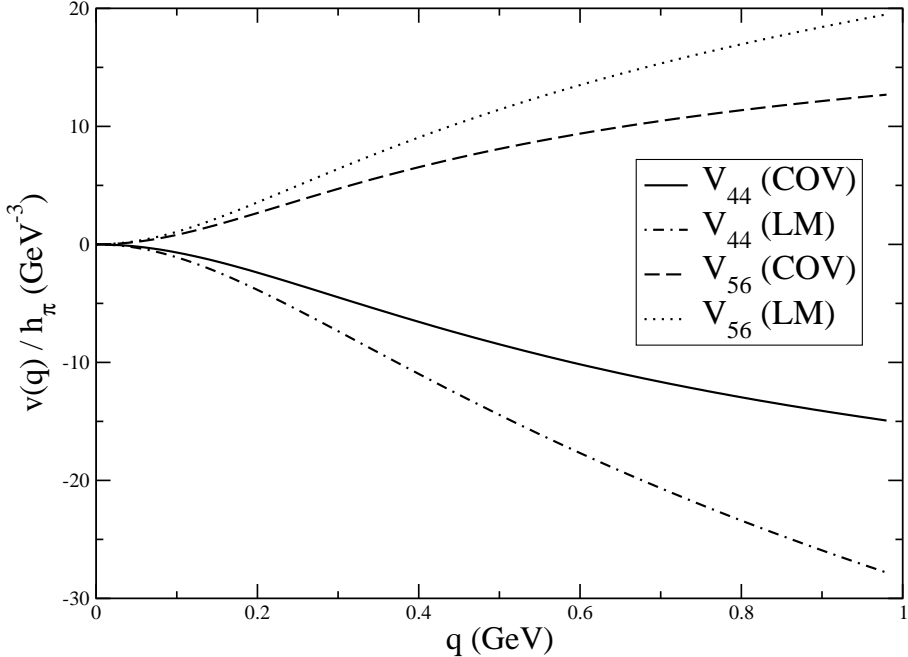}}
\end{center}
\caption{Subtracted potentials $v_{44}(q)$ and $v_{56}(q)$
for $q$ ranging from 0 to 1 GeV:  comparison of the large-$M$ limit approach
(LM, Eqs. (\ref{eq:vlargeM}))  with the covariant one. 
The asymptotic $q$ dependence of  the ``covariant" results is a constant one 
while the one for large-$M$ limit results has an extra ln$(q)$ dependence. 
\label{fig:su1000}}
\end{figure} 
\begin{figure}[htb]
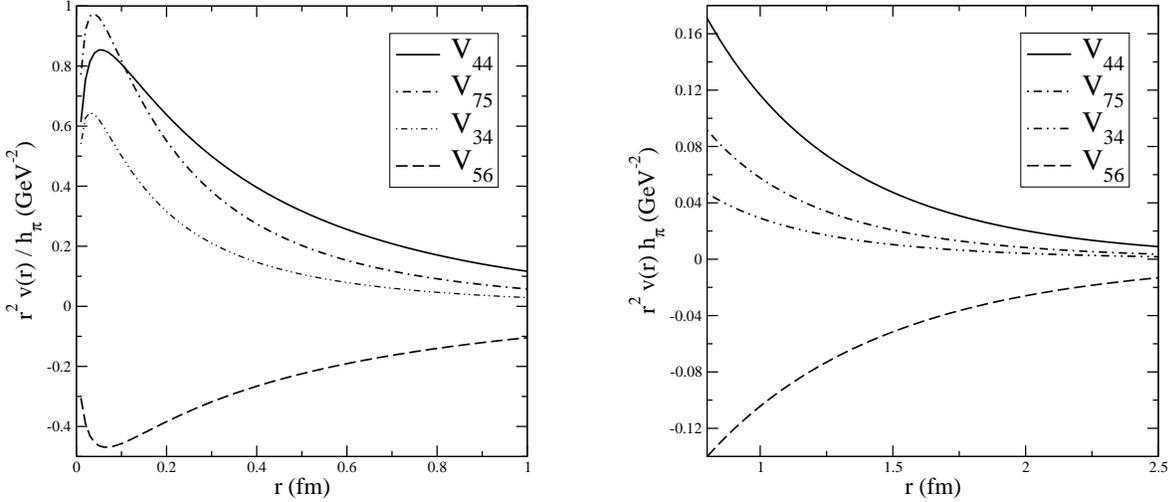

\begin{center}
\mbox{\epsfig{ file=potrfus10.eps, width=7cm}  \hspace*{1cm}
\epsfig{ file=potrfus25.eps, width=7.2cm} }
\end{center}
\caption{Potentials $v_{44}(r)$, $v_{34}(r)$, $v_{56}(r)$ and $v_{75}(r)$ 
at small and intermediate distances (left and right panels respectively). 
The results shown in the figure represent the above 
potentials multiplied by a phase factor, $r^2$ (in units of GeV$^{-2}$), 
to better emphasize the most relevant range for applications.  
\label{fig:vrs}}
\end{figure} 
\begin{figure}[htb]
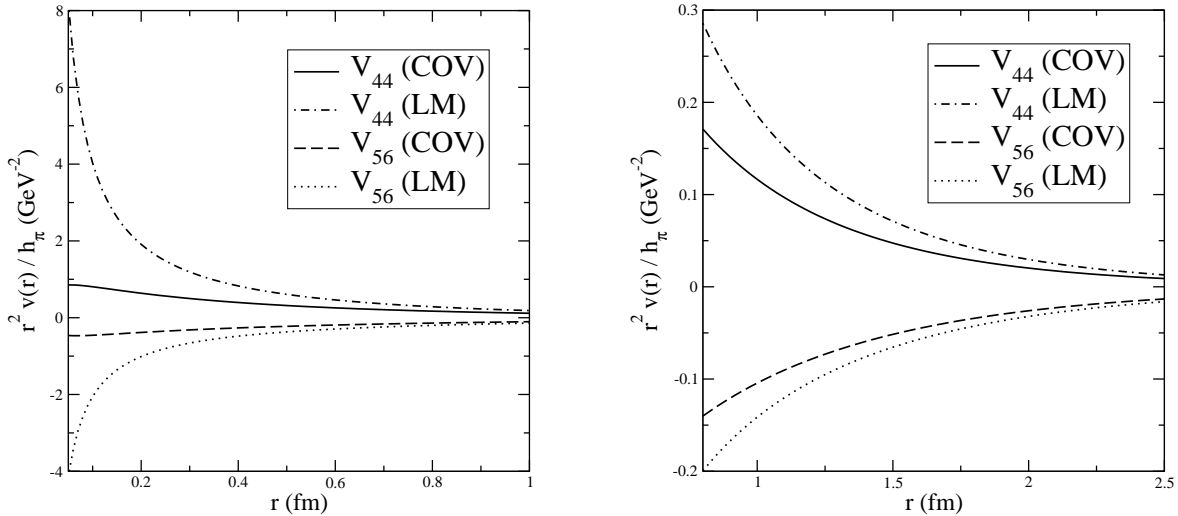

\begin{center}
\mbox{ \epsfig{ file=potrfugms10.eps, width=7cm} \hspace*{1cm}
\epsfig{ file=potrfugms25.eps, width=7.25cm}}
\end{center}
\caption{Potentials $v_{44}(r)$ and $v_{56}(r)$ at small 
and intermediate distances (left and right panels respectively): 
comparison of the ``covariant" calculations with their large-$M$ limits; 
other  definitions or comments as in Fig. \ref{fig:vrs}. 
The curves corresponding to the large-$M$ limit in the left panel 
tend to $\infty$ when $r\rightarrow 0$ and have a zero-range part, 
not shown in the figure, with an opposite sign. 
\label{fig:vrgm}}
\end{figure} 

We begin with the first four 
spectral functions $g_{ij}(t')$ entering potentials 
$V_{44}$, $V_{34}$, $V_{56}$ and $V_{75}$. Their $t'$ dependence 
is shown for a range of $t'^{1/2}$ going from threshold 
to about $5\,{\rm GeV}$
in Fig. \ref{fig:gts} (left panel). It is aimed to roughly evidence 
the relative weight of various components at small and high $t'$.
At first sight, different potentials have comparable sizes. 
The small-$t'$ range is physically more relevant in the sense 
that the regime beyond $1\,{\rm GeV}^2$ is expected to involve 
the contribution of other multi-meson exchanges. The higher-$t'$ range 
is more appropriate to illustrate convergence properties.  
At low  values of $t'$, one can see some significant differences 
as expected from the $1/M$ expansion, Eq. (\ref{eq:glargeM}). 
For $t'^{1/2}\leq 0.5\,{\rm GeV}$, the spectral functions entering 
the local potentials, $V_{44}$ and $V_{56}$, dominate those 
of the non-local ones, $V_{34}$ and $V_{75}$. 
All of them increase in the lower-$t'$ range 
(except in a very small $t'$ range for $V_{56}$)
and one has to go to much higher values of this variable to observe some
saturation and ultimately some decrease. The maximum is roughly reached 
around $t'^{1/2}\!=\!2M$, which, apart from the pion mass, 
is the only quantity entering the calculations.
The decrease, roughly given by $t'^{-1/2}$, up to log terms, ensures 
good convergence properties for potentials $v_{ij}(q)$,
Eq. (\ref{eq:disper}) (this would not be the case for the other 
Green function mentioned in the text, see details in Appendix \ref{app:green}).
The validity of the $1/M$ expansion for the spectral functions 
$g_{44}(t')$ and $g_{56}(t')$, which allows one to recover the EFT results 
for the essential part, can be checked by examining Fig. \ref{fig:gts} 
(right panel),  where the ``covariant" and approximate results 
are shown for a $t'$ range extending to $1\,{\rm GeV}^2$. 
It is observed that the dominant term in the $1/M$ expansion 
tends to overestimate the more complete
results both at very low and high values of $t'$. In the first case, 
the threshold behavior ($q_{\pi}^3$ for $g_{44}(t')$, arctg$(m^2_{\pi}/q_{\pi}M)$ 
for $g_{56}(t')$) is missed (see observation on the $1/M$ expansion 
in the previous section). 
In the second case, the overestimation, which is roughly given 
by a factor $t'/M^2$, tends to increase with $t'$, preventing one from getting
convergent results.  

In Fig. \ref{fig:vqs} (left panel), we show the potentials $v_{ij}(q)$ 
up to $q\!=\!1$ GeV (together with the OPE one that has a strong dependence 
on $q$ and is divided by 10 to fit the figure). 
As expected from examination of the spectral functions, 
the TPE potentials have roughly the same size and there is no strong evidence 
that some of them should be more important than other ones. 
It is also noticed that their decrease in the range $q=$(0$-$1) GeV 
is slower than  the standard $\rho$-exchange one, 
given by $1/(m_{\rho}^2\!+\!q^2)$, 
indicating they roughly correspond to a shorter-range interaction.
The comparison of the potentials, $v_{44}(q)$ and $v_{56}(q)$, 
with the corresponding EFT ones, Eqs. (\ref{eq:eft}),  
can be made by looking at Fig. \ref{fig:vqs} (right panel). 
The choice of $\Lambda_\chi$  in the EFT results 
($\Lambda_\chi=4\pi \,g_A\,M/g_{\pi NN}$)  is suggested by
the large-$M$ limit of the ``covariant" expression, see Eqs. (\ref{eq:vlargeM}),
to make the comparison as meaningful as possible. 
A striking feature appears here: the EFT potentials have a sign opposite 
to the ``covariant" ones but the $q$ dependence is roughly the same. 
This feature suggests that the LEC part could play an important role.

To make a more significant comparison, we subtracted a constant 
from both potentials so that they vanish at $q^2\!=\!0$. 
This procedure amounts to using subtracted dispersion relations, 
which leads to convergence in all cases. Moreover, in configuration space, 
this part of the interaction determines the long-range component 
of the potential, which is physically the most relevant one. 
Results are shown in Fig. \ref{fig:su1000}.
The ``covariant" and LM results have now the same sign. It is however noticed 
that the present LM results tend to overestimate the ``covariant" ones. 
In the limit $q\rightarrow0$, the overestimate reaches a factor 1.6 for $V_{44}$ 
and a factor 1.3 for $V_{56}$. This points in this case 
to the role of higher $1/M$-order corrections. Actually, 
the discrepancy vanishes in the limit $m_{\pi}/M \rightarrow 0$, 
showing that the non-zero pion mass has some effect. The discrepancy 
tends to slowly increase with $q$ (factors 1.9 and 1.5 at $q \!=\!$1 GeV 
for $V_{44}$ and $V_{56}$ respectively), pointing out to the role 
of ln($q$) corrections appearing in the large-$M$ limit. Altogether, 
these results are in accordance with the overestimate already noticed 
for spectral functions in this approximation. 
>From a different viewpoint, these results confirm the expectation 
that the discrepancy between the EFT and ``covariant" approaches 
shown in the right panel of Fig. \ref{fig:vqs} can be ascribed to contact
terms. A rough agreement would be obtained with the effective potential 
obtained in the $\overline{ \mbox{\rm MS}} $ together 
with a dimensional-regularization scale $\mu$ ranging 
from $3 \,m_{\pi}$ to $6\, m_{\pi}$, depending on how this is made 
(see the definition of this scheme at the end of Appendix \ref{app:eft}).

The Fourier transforms of the $v_{ij}(q)$ quantities,  $v_{ij}(r)$, are shown 
in Fig. \ref{fig:vrs} for small distances (left panel) 
as well as intermediate distances (right panel). We call them potentials 
though they are dimensionless quantities, the energy
dimension being given by the extra operators $\vec{p}$.
They are multiplied by a factor $r^2$ to emphasize 
the range which is relevant in practice for calculations. 
At small distances, the comparison of
various components roughly reflects the one for spectral functions or
potentials in momentum space. Examination of these results at large distances
evidences some significant differences. They have a better agreement 
with expectations from the $1/M$ expansion or from the very-low-$t'$ 
behavior of spectral
functions. Thus, the local potentials, $v_{44}(r)$ and $v_{56}(r)$, 
have a range larger than the other two components, 
$v_{34}(r)$ and $v_{75}(r)$, do. The differences appear only in the range 
where potentials have  small contributions to PV effects. 

The comparison with the LM potentials is given in Fig.~\ref{fig:vrgm} 
for the local components ($V_{44}$ and $V_{56}$). 
It is noticed that these last potentials should be completed 
by contact terms, which  have a sign opposite to the corresponding curves.
Due to the difficulty of representing these terms in a simple way, 
they have not been drawn in this figure. Moreover, they are not 
distinguishable from the LEC's contributions and have thus an arbitrary
character (they depend on the subtraction scheme). 
Considering first potentials at intermediate (or long) distances, 
where they can be the most reliably determined, it is found
that the LM and ``covariant" potentials have the same sign despite they 
have opposite sign in momentum space. This result is in complete accordance 
with the  subtracted potentials shown in Fig. \ref{fig:su1000}. 
Indeed, the slope of the corresponding results is, up to a minus sign, 
a direct measure of the square radius of the potential weighted 
by its strength. The negative slope for the potential $V_{44}$ 
thus indicates that its configuration-space representation 
is positive at intermediate distances (the opposite for $V_{56}$). 
Quantitatively, the significant dominance of the LM results over 
the ``covariant" ones (factors 1.7 and 1.4 for $V_{44}$ and  $V_{56}$ 
respectively at $r=0.8$ fm) confirms what is found 
for subtracted potentials in momentum space. 
Considering now the very short-range domain, it is found that the product 
of the LM potentials with $r^2$, which are shown in Fig. \ref{fig:vrgm} 
(left panel), diverge like $1/r$ when $r \rightarrow 0$, up to log factors. 
The contribution of this part alone to the plane-wave Born amplitude is thus
logarithmically divergent. This contribution turns out to be canceled 
by the zero-range one, mentioned above, so that the sum is finite.
Thus, there is no principle difficulty with this peculiar behavior
of the LM potentials in configuration space but, of course, some care is
required in estimating their contribution.

\begin{figure}[htb]
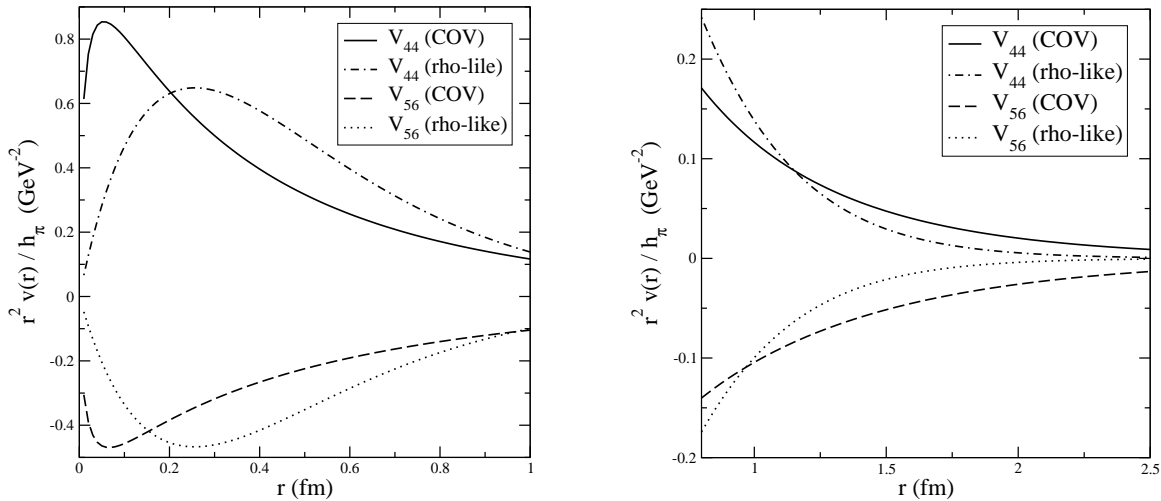

\begin{center}
\mbox{ \epsfig{ file=potrfuros10.eps, width=7.0cm} \hspace*{1cm}
\epsfig{ file=potrfuros25.eps, width=7.05cm}}
\end{center}
\caption{Potentials $v_{44}(r)$ and $v_{56}(r)$  
at small and intermediate distances (left and right panels respectively): 
comparison of the ``covariant" calculations with a standard rho-exchange potential 
normalized to the same volume integral; 
other definitions or comments as in Fig. \ref{fig:vrs}.
\label{fig:rho}}
\end{figure} 

Naively, it could be thought that the TPE is a medium-range interaction, as
already mentioned. In Fig. \ref{fig:rho}, we compare potentials 
$v_{44}(r)$ and $v_{56}(r)$ to a standard rho-exchange one normalized 
so that they have the same volume integral. 
This quantity determines the low-energy plane-wave Born amplitude 
(up to a factor $\vec{p}$ which can be factored out). 
At very large distances, the effect of the longer-range TPE tail 
is evident but this occurs in a domain where the potential 
is quite small and will not contribute much. At intermediate 
or even at small distances however,  
the TPE roughly compares to the  $\rho$ exchange. 
Actually, it turns out to have a shorter range. This reflects 
the fact that the two-pion continuum has an unlimited mass 
(the integration over $t'$ extends to infinity). 
Moreover, it is slightly more singular at very small distances, 
as a result of the extra ln(r) dependence of the TPE potential. 
This last effect is typical of relativistic effects. 
Both long- and short-distance effects can be traced back 
to the $g(t')$ function which, in the TPE case, extends to both small 
and large values of $t'$ with a maximum around $t'\!=\!4M^2$, 
while it is concentrated around $t'\!=\!m^2_{\rho}$  
for the $\rho$-exchange one ($\delta(t'\!-\!m^2_{\rho})$ function 
in the zero-width limit).

For simplicity, we did not consider explicitly the contribution due to
nucleon resonances in the two-pion box diagrams, which was accounted for 
in the 70's in a ``covariant" approach \cite{chem_npa74} 
or recently in a EFT  \cite{kaiser_prc07,liu_arxiv07} 
or a TO one \cite{niskanen_arxiv07}. As already mentioned, 
part of it is contained in the EFT approach by relying 
on the Weinberg-Tomozawa description of the $\pi N$ scattering amplitude.
It only contributes to the $V_{56}$ component and could represent 
10$-$20\% of the total contribution, depending on the range 
and how it is estimated. 
On top of it, there are further contributions which affect 
both $V_{44}$ and $V_{56}$ components. In the case of  $V_{44}$, 
they amount to an extra 30$-$40\% contribution  around 1 fm 
in the ``covariant" calculation \cite{chem_npa74} 
and roughly twice as much in the EFT approach \cite{kaiser_prc07,liu_arxiv07} 
or the TO one \cite{niskanen_arxiv07}. 
For the case of  $V_{56}$, it is more complicated 
as the above mentioned contribution relative to the description 
of the  $\pi N$ scattering amplitude has to be disentangled first 
for the ``covariant" calculation. When this is done, the extra contribution 
due to resonances decreases and could represent 30$-$40\% 
of the contribution with nucleons only \cite{chem_npa74}. 
This is slightly less than what is obtained in the EFT approach 
\cite{kaiser_prc07}. Taking into account that the EFT results 
overestimate the ``covariant" ones for the nucleon intermediate state,
it thus appears that the overestimate for the resonance intermediate state 
is significantly larger. This feature points to corrections 
of order $p^2/(M(M_{\Delta}\!-\!M))$ affecting the resonance propagator 
(dispersion effects), which are known to be important, 
while corrections in relation with the large-$M$ limit are 
of order $p^2/M^2$. It reinforces the conclusion of Ref.  \cite{kaiser_prc07}
that, contrary to the strong-interaction case, the role of resonances, 
especially the $\Delta(1232 \;{\rm MeV} )$ one, plays a negligible role 
in the PV $NN$ interaction. 

\section{Estimates of PV effects in two processes}
The TPE potentials considered here have an isovector character. 
At low energy, they can contribute to two different transitions,
$^1S_0\!-\!\!\,^3P_0$,  which involves identical particles 
like two protons or two neutrons and $^3S_1\!-\!\!\,^3P_1$, 
which involves different particles, proton and neutron. 
The interactions $V_{44}$ and  $V_{34}$ contribute in the first case 
while the interactions $V_{56}$ and  $V_{75}$ contribute in the other one.
Two processes of current interest, where the TPE potentials matter, 
are respectively proton-proton scattering and radiative neutron-proton 
capture at thermal energy. In the first case, a helicity dependence 
of the cross section, $A_L(E)$, has been measured at different energies 
\cite{kist_prl87,ever_plb91,berdoz_prl01}. 
In the second case, an asymmetry in the direction of the photon emission 
with respect to the neutron polarization, $A_{\gamma}$, has been looked for 
at LANSCE \cite{lansce} (the experiment is now running at SNS). 
The TPE contribution to these effects is discussed in the following. 
The calculations have been performed with the $NN$-strong-interaction model, 
Av18 \cite{wir_prc95}, which is
local wave by wave. Vertex form factors are ignored. 
On the one hand, the dispersion-relation formalism assumes on-mass-shell
particles, the contribution due to form factors in other approaches being
generated  by what is included in the dispersion relations. 
On the other hand, the role of form factors was already examined 
within the EFT approach \cite{hyun_plb07,liu_xxx06}, partly 
with the motivation of regularizing 
a potential that is badly behaved at short distances, 
$\propto (r^{-3}\!-\! c\,\delta(\vec{r}\,))$ where $c$ is infinite and  ``determined" 
so that the integral over $\vec{r}$ has a well-defined value. There is no
principle difficulty to work with this potential, however,\footnote{The trick is
to separate in the integrands a part determined by wave functions at the
origin, of which the integral over $\vec{r}$ is known, the remaining part being
well behaved at the origin.}
and we will therefore use it here. This will facilitate 
the comparison with the ``covariant" results. 

As a side remark, we notice that form factors different from those mentioned
above have been used in the context of applying effective field theories
to the strong $NN$  interaction \cite{epel_05}. 
They involve a separable dependence of the
relative momentum in the initial and final states, $\vec{p}$ and $\vec{p}\,'$,
instead of $\vec{q}=\vec{p}-\vec{p}\,'$. Their effect is to smooth out wave
functions at short distances in accordance with the idea that the corresponding
physics, partly unknown, should be integrated out and accounted for by LEC's. 
For such form factors, it would be more convenient to work in momentum space.
However, in the case where configuration space is chosen, the methods we used 
for dealing with the badly-behaved potential could be useful there too. 
\subsection{Proton-proton scattering}
The first calculation of TPE effects was done 
by Simonius \cite{sim_plb72}, with the aim to get some estimate 
for a measurement of PV effects in $pp$ scattering (the Cabibbo model 
then used was not contributing to the $pp$ force in its simplest form). 
Our results, obtained here for three energies at which the longitudinal
asymmetry $A_L(E)$ has been measured, are presented in Table \ref{tab:A_L}.

Examining the ``covariant" results, it is found that the contribution 
of the local term, $V_{44}({\rm COV})$, dominates over the non-local one,
$V_{34}({\rm COV})$, which appears at the next order in the $1/M$ expansion. 
The result could be guessed from looking at Fig. \ref{fig:vrs}.
Their ratio is of the order of  the factor $1+\mu_V=4.706$ which appears 
in the $\rho$-exchange potential. There are reasons to think this result is not
accidental (see end of Sect. \ref{sec_relation}). 
The present results compare to the earlier ones \cite{sim_plb72} 
as well as with the value  $A_L(13.6\, {\rm MeV})=-0.1\, h^1_{\pi}$
that has been used in analyses of PV effects \cite{desp_phrep98}. 
The closeness can be attributed to the fact that the Av18 model employed here
and the Reid or Hamada-Johnston models previously used are local ones and,
moreover, evidence a strong short-range repulsion. 
Significant departures could occur, instead, with models 
like CD-Bonn \cite{mach_prc01} or some Nijmegen ones \cite{stoks_prc94}, 
which have a non-local character (see Ref. \cite{amghar2_npa03} for a discussion
about the role of non-locality). 
\begin{table}[hb]
\caption{PV asymmetries in $pp$ scattering at three energies, 
13.6, 45 and 221 MeV: 
successively for the potentials $V_{44}({\rm COV})$, $V_{34}({\rm COV})$, 
$V_{44}({\rm rho}$-${\rm like})$ and  $V_{44}({\rm LM})$, in units 
of the $h^1_{\pi}$ coupling constant.\label{tab:A_L}\vspace*{0.2cm}}
\begin{tabular}{lccc}
 \hline  
  Energy (MeV)            &   13.6   &   45     &  221    \\  [1.ex] \hline
 $V_{44}({\rm COV})$      & -0.092   & -0.154   & 0.072   \\  [0.ex] 
 $V_{34}({\rm COV})$      & -0.022   & -0.042   & -0.029  \\  [0.ex]
$V_{44}({\rm rho}$-${\rm like})$ & -0.110   & -0.196   & 0.096   \\  [0.ex] 
 $V_{44}({\rm LM})$       & -0.150   & -0.252   &  0.127  \\  [0.ex] \hline     
\end{tabular}
\end{table}
 
Due to its long-range component, one could infer that the TPE contribution
should be enhanced with respect to the $\rho$-exchange one when the effect of 
short-range repulsion is accounted for. The comparison 
of $V_{44}({\rm rho}$-${\rm like})$ and $V_{44}({\rm COV})$ results shows that 
this is the other way round. This can be explained by the fact that
the long-range contribution where the TPE dominates over the $\rho$-exchange
has  little contribution to the asymmetry (a few \%). 
Instead, the short-range contribution where  the TPE dominates 
over the $\rho$-exchange plays a bigger role. The effect of short-range
repulsion on this contribution will therefore be enhanced, hence decreasing 
of the total TPE contribution with respect to  the $\rho$-exchange one.

\subsection{Asymmetry in neutron-proton radiative capture}
Contrary to $pp$ scattering discussed above, the asymmetry  $A_{\gamma}$
 involves a non-zero contribution from OPE.  
Thus, independently of the value of the coupling $h^1_{\pi}$, one can 
directly compare the TPE and OPE contributions.  
The last one has been extensively studied, 
see Ref. \cite{desp_plb01} and earlier references therein, 
and Refs. \cite{hyun_plb07,liu_xxx06,hyun_plb01} for more recent works. 
Its contribution is approximately given by $A_{\gamma}(OPE)= -0.11 h^1_{\pi}$ 
($= -0.112 h^1_{\pi}$ for the Av18 model used here). 

The TPE contributions to  $A_{\gamma}$ for the ``covariant" case are given by:
\begin{eqnarray}
&&A_{\gamma}(V_{56}({\rm COV}))= 0.0093 \,h^1_{\pi}, 
\nonumber \\
&&A_{\gamma}(V_{75}({\rm COV}))= - 0.0040\, h^1_{\pi},
\end{eqnarray}
while those for the rho-like  and large-$M$ limit are:
\begin{eqnarray}
&&A_{\gamma}(V_{56}({\rm rho\!-\!like}))= 0.0093 \,h^1_{\pi}, 
 \\
&&A_{\gamma}(V_{56}({\rm LM}))= 0.0141\, h^1_{\pi}.
\end{eqnarray}
Considering the ``covariant" results, it is first noticed that the contribution
of the local term, $V_{56}({\rm COV})$, dominates over the non-local one,  
$V_{75}({\rm COV})$, which appears at the next $1/M$ order. 
The effect is however less important than in $pp$ scattering, 
which can be inferred from looking at the corresponding potentials 
in Fig. \ref{fig:vrs}. 
Moreover, contrary to this process, their contributions have opposite signs. 
Thus, the total contribution represents only $-$5\% of the OPE one.
This is good news in the sense that it confirms that the asymmetry  
$A_{\gamma}$ represents the best observable to determine 
the coupling $h^1_{\pi}$. 
Present results reasonably compare to an earlier one \cite{desp_npa75}: 
$A_{\gamma}(TPE)=0.008\,h^1_{\pi}$, which is added up by
$A_{\gamma}(V_{56})= 0.0107 \,h^1_{\pi}$ and  
$A_{\gamma}(V_{75})=-0.0027\, h^1_{\pi}$.
Part of the difference for $A_{\gamma}(V_{56})$ can be traced back to the 
omission here of the $\Delta$ resonance contribution in intermediate states. 
It is reminded that this contribution tends to make the $\pi N$-scattering 
amplitude consistent with the  Weinberg-Tomozawa coupling, resulting 
in an enhancement of
the $V_{56}$ interaction (see discussion in Sec. \ref{sec_relation}). 
The main difference for  $A_{\gamma}(V_{75})$ is due to the extension 
of the integration over $t'$ in the dispersion relation from $50\,m_{\pi}^2$ 
to $\infty$ here.  

At first sight, the comparison of the TPE and $\rho$-exchange results, 
which  are essentially the same, shows features different from  
those observed in $pp$ scattering. Examination of Fig. \ref{fig:rho} 
indicates that, in comparison to the $V_{44}$ potential,  
the enhancement of the TPE potential, $V_{56}({\rm COV})$, 
over the $\rho$-exchange one, $V_{56}({\rm rho}$-${\rm like})$,  
is larger at large distances and smaller at short distances. 
As a result, the two effects due to the short-range repulsion 
mentioned for  $pp$ scattering tend to cancel here.

\subsection{Results in the large-$M$ limit}
Involving the subtraction of an infinite term, results obtained 
in the large-$M$ limit, which can be identified with the EFT ones for the
essential part, cannot be directly compared to the ``covariant" ones.  
A first look nevertheless shows that the results, apart from an enhancement 
by a factor 1.5 or so, are very similar. Moreover, from considering the
plane-wave Born approximation, we should have expected opposite signs. It is
therefore appropriate to give some explanation helping to understand these
results.

We first checked which range was contributing to PV asymmetries and found that
the role of the range below 0.4$-$0.5 fm was rather small (10$-$20\%). 
The observed enhancement of the large-$M$ result over the ``covariant" one 
thus reflects the similar enhancement that can be inferred 
from the corresponding potentials in Fig. \ref{fig:vrgm}, around 0.8 fm. 
The little role of the range below 0.4$-$0.5 fm is not expected 
from considering Fig. \ref{fig:vrs}. Instead, it can
be related to the strong interaction model, Av18, used to calculate 
wave functions entering estimates of observables. This model evidences the
effect of a strong repulsion at short distances. As this effect is much less
pronounced with non-local models, one can expect that the use of models like
CD-Bonn or some Nijmegen ones, will show some differences with
the present results. We can anticipate an  increase of the magnitude 
of the ``covariant" results and a  decrease of the LM ones (without excluding 
in this case a change in sign reminding that one for the plane-wave Born
approximation).

While trying to understand the role of higher $1/M$-order corrections, 
we face the problem that the corresponding contributions to potentials 
are more singular than those for $V_{44}({\rm LM})$ or $V_{56}({\rm LM})$, 
requiring some regularization and introduction of further LEC's. 
Restricting our study to the range $r \geq$ 0.4$-$0.5 fm, which provides most of
the contribution to the PV asymmetries, we found that the contribution at the
next $1/M$ order had a sign opposite to the dominant one, which it largely
cancels in the range around 0.8 fm. The correction, which is larger than needed, 
suggests that other corrections are necessary to ensure
reasonable convergence. Among them, one could involve 
chiral symmetry breaking.  We checked that in the limit of a massless pion, 
the ``covariant" potential $v_{44}(r)$ and its large-$M$ limit 
are significantly closer to each other. A non-zero pion mass 
could thus explain a sizable part of the discrepancy for observables 
between the ``covariant" and the large-$M$-limit results.
We notice that the comparison of the EFT and ``covariant" approaches 
in the strong interaction case evidences features similar 
to the above ones (see Ref. \cite{higa06} and references therein). 
As far as we can see, the similarity is founded for a part.

The large-$M$ limit of ``covariant" potentials assumes a particular
subtraction scheme. In another scheme, the $\overline{\mbox{\rm MS}}$  one, 
part of the functions $L(q)$ appearing in Eqs. (\ref{eq:vlargeM}) are replaced 
by  $L(q)\!-\!1\!-\!{\rm ln}(\mu/m_{\pi})$ (see Appendix \ref{app:eft}). 
Due to the effect of short-range repulsion in the Av18  
model, the correction has little effect on the calculated observables 
(a few \% increase for the $S\!-\!P$ transition part). 
This would be different for non-local models where the correction could be
significantly larger. Depending on the value of the parameter $\mu$, a large
part of the sensitivity of the LM results to strong-interaction models
mentioned above could be removed \cite{hyun_plb07}.  We notice that 
a discussion similar to the above one on the role of the subtraction scheme 
has also been held in the strong-interaction case \cite{epel_05}. 
It was concluded that the spectral function regularization 
scheme (SFR) was providing better convergence properties 
than for the dimensional-regularization one, due to avoiding spurious 
short-range contributions. To some extent, the SFR scheme is close 
to the $\overline{\mbox{\rm MS}}$ one considered here, 
with the parameter $\tilde{\Lambda}$ 
introduced in the former case being replaced by the quantity 
$2\mu$ in the latter case. 
%
\section{Conclusion}
In the present work, we have compared different approaches for incorporating 
the two-pion-exchange contribution to the PV $NN$ interaction. 
They include a covariant one, which fully converges and can thus be 
considered as a benchmark, and an effective-field-theory and 
a time-ordered one which can contain infinities. The last two approaches 
involve two components at the leading order, with a local character, 
while the covariant one involves both local and non-local. 
For a given transition, one can thus compare the local components 
obtained from different approaches on the one hand, local and non-local 
components on the other hand. These two comparisons can allow one 
to assess how good is the assumption of dominant order 
in the effective-field theory approaches as well as the role ascribed to LEC's. 

We first notice that  the effective-field theory (EFT) approach, 
the time-ordered-diagram approach  
and the limit of the covariant one at the lowest non-zero order 
in the $1/M$ expansion (LM) essentially agree 
with each other for the local terms. 
Possible discrepancies involve ingredients that have been omitted 
(contribution of baryon resonances in parti\-cu\-lar)
but are unimportant for the comparison. We can thus concentrate 
on a comparison of the covariant approach with its LM limit. 
Taking into account that this approach is determined  up to contact terms, 
rough agreement is found.
This is better seen by considering the subtracted potentials in
momentum space or intermediate distances ($r=1$fm) in configuration space. 
Quantitatively, the LM (EFT) approach tends to overestimate 
the ``covariant" results. At low $q$ or at intermediate distances, 
the effect reaches factors 1.7 and 1.3 respectively 
for the transitions $^1S_0\!-\!\!\,^3P_0$ and  $^3S_1\!-\!\!\,^3P_1$. 
At very small but finite distances, 
the LM (EFT)  potentials become very singular and their contribution to
physical processes diverge. This divergence is canceled 
by the contribution associated to the contact term so that the total result 
is finite (after renormalization). This peculiar behavior at $r=0$ and 
around contrasts with the smooth but diverging behavior in momentum
space of the published EFT two-pion-exchange interaction. In this case, 
it turns out that the sign of the potential is opposite to that one 
at finite distances, which is the most relevant part. 
This suggests that this EFT two-pion-exchange potential is
dominated by an unknown contact term, as far as a comparison with the 
``covariant" result is concerned. The problem disappears with a different
subtraction scheme, like the minimal one with a dimensional-regularization
scale $\mu$ in the range (3$-$6)$\, m_{\pi}$. This last choice tends to minimize 
the role of short distances, confirming the absence 
of a large sensitivity  to cutoffs observed elsewhere \cite{hyun_plb07}.
Interestingly, the $\overline{\mbox{\rm MS}}$ scheme corresponds to cutting off 
the dispersion integrals at a value of $t'^{1/2}$ around $2\mu$,  
which, together with the above value of $\mu$, 
is about (1$-$2)\,GeV. 
This is quite a reasonable value for separating the contribution 
of known physics from the unknown one to be integrated out. 
We thus believe that the choice of the  $\overline{\mbox{\rm MS}}$ scheme would be 
more appropriate, the interaction then ascribed 
to the EFT two-pion-exchange one being a better representation 
of the most reliable part of the two-pion-exchange physics,
which occurs at intermediate and large distances. 

Comparing the non-local components to the local ones, it is found that 
they are rather suppressed at large distances. For such distances, the
contribution to dispersion relations is expected to come from values of $t'$
smaller than $M^2$. The suppression of non-local terms then reflects the fact
that they have an extra $1/M$ factor in the $1/M$ expansion. At short distances, 
instead, the contribution to dispersion relations comes from large values of
$t'$. In this case, the local and non-local terms have the same $1/M$ order 
and they tend to have comparable contributions. Another aspect of 
the dependence on the range concerns the comparison with the $\rho$-meson
exchange. Not surprisingly, the two-pion exchange dominates the $\rho$-meson 
exchange one at long distances but, the potential being relatively small there, 
not much effect is expected from this part on the calculation of observables. 
The two-pion-exchange contribution is slightly dominated by the $\rho$-meson
exchange one at medium distances and dominates again the last one at very short
distances. Taking into account that the dominant contributions come from short
and intermediate distances, the two-pion-exchange contribution turns out to have
a shorter range than the $\rho$-meson exchange.

We looked at the TPE contribution in two physical processes, 
$pp$ scattering and radiative thermal neutron-proton capture.  
Roughly, they confirm what could be inferred from examining potentials. 
The results from the ``covariant" approach, calculated 
with the Av18 $NN$ strong-interaction model, 
essentially agree with earlier estimates based on other models.  
The main discrepancies evidence the role of some inputs such as 
the restriction on the $t'$ value in the dispersion relations or
the role of resonances in modeling the $\pi N$ scattering amplitude 
which enters these relations and we omitted here for simplicity.
Despite plane-wave Born amplitudes calculated 
with the EFT two-pion-exchange and the ``covariant" approaches
have opposite signs, it turns out that their contributions 
to observables are relatively close to each other. 
This feature points to the Av18 model, which produces 
wave functions that evidence the effect of a strong repulsion at
short distances. Such a property is interesting in that 
the main contribution to observables comes from intermediate and
long distances, where the derivation of the TPE potential is the most reliable. 
It is thus found that the EFT two-pion-exchange at NNLO tends 
to overestimate the ``covariant" results by about 50\%, 
pointing to a non-negligible role of next order corrections. 
This is confirmed by the consideration of non-local terms, 
which correspond to higher-order terms. Their contribution 
is especially important in neutron-proton capture, 
due to a destructive interference with the dominant one. 
Thus, the result at the dominant $1/M$ order (LM) in this process 
exceeds the ``covariant" one by a factor of about 2$-$3.
Interestingly, present results are rather insensitive to the subtraction scheme
as far as the coefficient, $1\!+\!{\rm ln}(\mu/m_{\pi})$, remains in the range 
of a few units. This is a consequence of the strong short-range repulsion 
present in the Av18 model. We however stress the fortunate character of
this result. The use of non-local strong-interaction models, like CD-Bonn 
or some Nijmegen ones, could lead to different conclusions. 
Actually, far to be a problem, the dependences on the model expected 
for the contribution of the term  $1\!+\!{\rm ln}(\mu/m_{\pi})$ 
and the EFT potential calculated in the maximal-subtraction (MX) scheme 
are likely to cancel for a large part. 

By comparing different approaches to the description of the PV TPE $NN$ 
interaction, it was expected one could learn about their respective relevance. 
Implying a natural cutoff of the order of the nucleon mass, 
the ``covariant" approach provides an unavoidable benchmark. 
This is of interest for the EFT approach which, up to now, 
has been considered at the lowest $1/M$ order. 
In improving this approach, a first step concerns  
the subtraction scheme. The minimal-subtraction one 
($\overline{ \mbox{\rm MS}}$), 
which involves a renormalization scale, $\mu$, 
is probably more favorable. By taking for this scale a value 
of the order of the nucleon mass (or the chiral-symmetry-breaking 
scale $\Lambda_{\chi}$), the scheme better matches the separation 
of the interaction into known and less known contributions, 
corresponding respectively to long  and short distances. 
The LEC's so obtained could be less dependent on
the strong-interaction model. 
The next step  should concern higher $1/M$-order terms, 
whose contributions are not negligible. 
This is likely  to require a lot of work and caution, as the singular behavior 
of these terms at short distances increases with their order.
Meanwhile, the ``covariant" results could provide 
both a useful estimate and a relevant guide for their study.

{\bf Acknowledgments}
We thank the Institute for Nuclear Theory at the University of Washington 
for its hospitality and the Department of Energy for partial support 
during the completion of this work.
The work of CHH was supported by the Korea Research Foundation Grant 
funded by the Korean Government (MOEHRD, Basic Research Promotion Fund) 
(KRF-2007-313-C00178).
The work of SA was supported by the Korea Research Foundation 
and the Korean Federation of Science and Technology funded 
by the Korean Government (MOEHRD, Basic Research Promotion Fund) 
and SFTC grant number PP/F000488/1.
The work of CPL was supported in part by the U.S. Department of Energy under
contract DE-AC52-06NA25396. 
We are grateful to Prof. U. van Kolck for pointing out an inconsistency of the
simplest time-ordered-diagram approach with chiral-symmetry expectations.

\appendix

\section{Subtraction of the iterated OPE and related questions}
\subsection{Expressions of the spectral functions, $g(t')$}
\label{app:gts}
Historically, the derivation of the isovector PV two-pion exchange started 
with the calculation of the crossed diagram \cite{pign_plb71}. 
The calculation of the non-crossed diagram, which implies the removing 
of the iterated OPE contribution and thus requires more care, 
came slightly later \cite{desp_plb72}.\footnote{The paper contains editor
mistakes that could obscure its understanding: the contents 
of figures 1 and 2 should be interchanged and the number -1.61 in the table
should be replaced by -0.61.} 
Last works along the same lines \cite{pirner_plb73,chem_npa74} 
considered the two types of diagrams on the same footing. 
We first remind here some results relative to the crossed 
and non-crossed diagrams with the notations 
of Ref. \cite{desp_plb72} (functions $g_{A}(t')$, $g_{B}(t')$, $g_{C}(t')$ 
and $g_{D}(t')$, $g_{E}(t')$ respectively). The functions $g_{A}(t')$ 
and $g_{D}(t')$ correspond to the same spin-isospin structure.
The two versions of the iterated OPE discussed in the text, 
which concern the $g_{D}(t')$ and $g_{E}(t')$ functions, 
are considered. The one employed in Ref. \cite{desp_plb72} 
corresponds to the term with the factor $(E+M)$ in the integrand  
while the other one considered in later works contains the factor $2E$. 
The expressions read:
\begin{eqnarray}
g_{A}(t') &=& 
\frac{1}{2M} \Bigg (\frac{G}{x}-\frac{H\;x}{x^2+4M^2q_{\pi}^2}\Bigg )\, ,
\nonumber \\
g_{B}(t') &=& \frac{x}{M^2} \;g_{A}(t')\, ,
\nonumber \\
g_{C}(t') &=& \frac{1}{2M} 
\Bigg (\frac{4q_{\pi}}{\chi^2}+\frac{H}{M^2} - 
G \Big (\frac{1}{M^2}+\frac{1}{\chi^2} \Big ) \Bigg)\, ,
\nonumber \\
g_{D}(t') &=& 
-\frac{x}{M^2 \; m^2_{\pi}}\; {\rm arctg} \Big (\frac{m^2_{\pi}}{2Mq_{\pi}}\Big )
-\frac{G}{2xM}
\nonumber \\
& & +\int_{k_{-}^2}^{{k_{+}^2}} 
\frac{dk^2}{k^2\sqrt{k^2t'\!-\!(m^2_{\pi}\!+\!k^2)^2}}
\Bigg ( \frac{2E\,{\rm or }\;(E\!+\!M)}{E^2}
\Big ( \frac{k^2\!-\!x}{4M}\!-\!\frac{E\!-\!M}{2}\Big )
\!+\! \frac{x}{2M^2}  \Bigg )\,,
\nonumber \\
g_{E}(t') &=&\frac{4x^2}{M^2 \, m^2_{\pi}\,t'}\;
 {\rm arctg} \Big (\frac{m^2_{\pi}}{2Mq_{\pi}}\Big )+\frac{2G}{Mt'}
\nonumber \\
& &\!\! +\int_{k_{-}^2}^{{k_{+}^2}} 
\frac{dk^2}{k^2\sqrt{k^2t'\!-\!(m^2_{\pi}\!+\!k^2)^2}}
\nonumber \\
& &\times \Bigg ( \frac{2E\,{\rm or }\;(E\!+\!M)}{E^2}
\Big ( \frac{(k^2\!-\!x)^2}{Mt'}\!-\!\frac{2(E\!-\!M)(k^2\!-\!x)}{t'}
\!+\!\frac{(E\!-\!M)^2}{M}\Big )
\!-\! \frac{2x^2}{M^2t'}  \Bigg )\,,
\nonumber \\
\label{eq:origin}
\end{eqnarray}
where the various functions, $q_{\pi}, \;x, \;\chi, \; G,\;H   $, 
are given in the text, Eq. (\ref{eq:defs}). 
The writing slightly differs from Ref. \cite{desp_plb72}. 
No non-relativistic approximation is made for the integrands. 
Moreover, the original term, ${\rm arctg}(2Mq_{\pi}/m^2_{\pi})/m^2_{\pi}$, 
has been transformed into 
$(\pi/2-{\rm arctg}(m^2_{\pi}/2Mq_{\pi}))/m^2_{\pi}$ 
and the factor $\pi/2/m^2_{\pi}$ has been inserted in the
integral using the relation 
$\int_{k_{-}^2}^{{k_{+}^2}} 
dk^2/\Big(k^2\sqrt{k^2t'\!-\!(m^2_{\pi}\!+\!k^2)^2}\;\Big)=\pi/m^2_{\pi}$.
With this rearrangement, it can be checked that the integrand has no
singularity at $k^2=0$. 

The $g(t')$ functions considered in the present work are related 
to the above ones  by the relations:
\begin{eqnarray}
g_{44}(t') &=&\tilde{K}\,g_{C}(t')\,, 
\nonumber \\
g_{34}(t') &=& \tilde{K}\,g_{B}(t')\,, 
\nonumber \\
g_{56}(t')&=&\tilde{K}\,(g_{A}(t')+g_{D}(t'))\,,
\nonumber \\
g_{75}(t')&=&\tilde{K}\,g_{E}(t') \,,
\end{eqnarray}
where $\tilde{K}$ is an overall constant given in Eq. (\ref{eq:defs}).
An important point to note is that the contributions 
of the term $ G/2xM$ in $g_A(t')$ and $g_D(t')$,  which dominate 
at low energy, exactly cancel. 
This cancellation is important in restoring the crossing symmetry for pions,
a property that is fulfilled by the effective pion-nucleon interaction 
introduced in the EFT approach (triangle diagram).

\subsection{Asymptotic behavior of the $g(t')$ functions}
\label{app:green}
The asymptotic behavior of the spectral functions 
$g_{44}(t')$ and $g_{34}(t')$, which only involve the crossed-diagram
contribution, can be easily obtained from their expressions, 
Eqs. (\ref{eq:gcad}). The dominant term is of the order $1/(M\sqrt{t'})$ up to
some log factors, see Eq. (\ref{eq:larget}). 
The asymptotic behavior of the two other spectral 
functions, $g_{56}(t')$ and $g_{75}(t')$, is considerably more complicated. 
Their analytic part contains terms with the behavior 
$-\tilde{K}\sqrt{t'}/2M^3$ and $\tilde{K}\sqrt{t'}/M^3$ respectively 
while the integral part requires careful examination. 

The dominant term in the integral is given by the part proportional to $x$
and $x^2$ in the integrands of $g_{56}(t')$ and $g_{75}(t')$ respectively. 
As the integrands are the same up to a factor $-4x/t'$, it is sufficient to
consider the first case. Its contribution becomes:  
\begin{eqnarray}
I_{56}& =& -\tilde{K}x\int_{k_{-}^2}^{{k_{+}^2}} 
\frac{dk^2}{k^2\sqrt{k^2t'\!-\!(m_{\pi}^2\!+\!k^2)^2}}
\Bigg ( \frac{2E \;({\rm or}\; E+M)}{4M\,E^2}\!-\! \frac{1}{2M^2}  \Bigg )
\nonumber \\
& =&\tilde{K}x\int_{k_{-}^2}^{{k_{+}^2}} 
\frac{dk^2}{k^2\sqrt{k^2t'\!-\!(m_{\pi}^2\!+\!k^2)^2}}
\Bigg (\frac {(E\!-\!M)}{2M^2\;E} \Big({\rm or}\; 
\frac {(E\!-\!M)(2E+M)}{4M^2\;E^2}\Big)\Bigg )
\nonumber \\
& =&\tilde{K}x\int_{k_{-}^2}^{{k_{+}^2}} 
\frac{dk^2}{\sqrt{k^2t'\!-\!(m_{\pi}^2\!+\!k^2)^2}}
\Bigg (\frac {1}{2M^2\,E\,(E\!+\!M)}\Big({\rm or}\; 
\frac {(2E+M)}{4M^2\;E^2(E\!+\!M)}\Big)\Bigg )
\, ,\label{eq:intv}
\end{eqnarray}
where the first case corresponds to the quadratic-energy dependent Green's
function retained here and the second case to the linear one.
After some algebra, it is found that in the large-$t'$ limit 
and a negligible pion mass, the integral with the first integrand writes:
\begin{eqnarray}
I_{56}(t' \rightarrow \infty) & \! \!\simeq \!\!&
-\frac{1}{2}I_{75}(t' \rightarrow \infty) \simeq  \frac{\tilde{K}}{M\sqrt{t'}}
\Bigg(\frac {t'}{2M^2}
-\frac{1}{8}\,{\rm ln} \Big(\frac{t'}{M^2}\Big) +\frac{1}{8}
-\frac{1}{2}\,{\rm ln} (2)\Bigg)\, .
\end{eqnarray}
The absence of the intermediate term $1/(M^2)$ results from a non-trivial
cancellation. All the other terms not considered here behave 
like $1/(M\sqrt{t'})$ (up to log terms) for the most important ones.
It can thus be checked that the contributions of order  $\sqrt{t'}/M^3$ 
to the spectral functions $g_{56}(t')$ and $g_{75}(t')$ cancel, leaving
contributions of order $1/(M\sqrt{t'})$. 

Have we used the other Green's function, the dominant contribution 
resulting from performing the integral would read:
\begin{eqnarray}
I_{56}(t' \rightarrow \infty)& \!\! \simeq \!\!&
-\frac{1}{2}I_{75}(t' \rightarrow \infty) \simeq 
\tilde{K}\frac {\sqrt{t'}}{2M^3}\;\frac{2+\pi}{4}\, ,
\end{eqnarray}
while the corresponding $g(t')$ functions would be given by: 
\begin{eqnarray}
g_{56}(t' \rightarrow \infty)& \!\! \simeq \!\!&
-\frac{1}{2}g_{75}(t' \rightarrow \infty) \simeq 
\tilde{K}\frac {\sqrt{t'}}{2M^3}\;\frac{\pi-2}{4}\, .
\end{eqnarray}
Thus, contrary to the quadratic-energy dependent Green's function, 
no cancellation of the do\-mi\-nant terms is found 
with the consequence that the dispersion integrals involving 
the spectral functions $g_{56}(t')$ and $g_{75}(t')$, Eqs. (\ref{eq:disper}),  
do not converge (logarithmic divergence). Moreover, these functions 
evidence a sign change  around $t'=$(20$-$25)$ \,{\rm GeV}^2$. 
Actually, this has little effect on the spectral  functions 
at low values of $t'$ but, of course, this prevents one from getting
convergent results. The sensitivity of the configuration-space 
potentials was not exceeding 10\% at $r=1$ fm \cite{chem_npa74}.

\section{$PV$ two-pion exchange $NN$ potential from EFT}
\label{app:eft}
In this Appendix, we give the detail of the momentum-space contributions 
from the two-pion exchange diagrams shown in Fig. \ref{fig:eft},  
employing heavy-baryon chiral Lagrangian and 
the dimensional regularization 
in the $d$-dimensional space-time, 
$d=4-2\epsilon$. 

>From the diagram (b), (c) and (d), we get:
\bea
V_{(b)} &\!=\!& 
-i(\vec{\tau}_1\times \vec{\tau}_2)^z
(\vec{\sigma}_1\!+\!\vec{\sigma}_2) \!\cdot\!\vec{q}\,
\frac{\pi g_Ah_\pi^1}{\sqrt2(4\pi f_\pi)^3}
\nnb \\ && \hspace*{2cm}\times
\left[
\frac{1}{\epsilon}
-\gamma 
+{\rm ln}(4\pi)
+{\rm ln}\left(\frac{\mu^2}{m_\pi^2}\right)
+2
-2L(q)
\right]\, , 
\\
V_{(c)} &\!=\!& 
-i(\vec{\tau}_1+\vec{\tau}_2)^z 
(\vec{\sigma}_1\times\vec{\sigma}_2) \!\cdot\! \vec{q}\, 
\frac{2\sqrt2\pi g_A^3h_\pi^1}{(4\pi f_\pi)^3}
\left[
\frac{1}{\epsilon}
-\gamma 
+{\rm ln}(4\pi)
+{\rm ln}\left(\frac{\mu^2}{m_\pi^2}\right)
+2
-2L(q)
\right]
\nnb \\ && - 
i(\vec{\tau}_1\times\vec{\tau}_2)^z
(\vec{\sigma}_1\!+\!\vec{\sigma}_2)\!\cdot\!\vec{q}\,
\frac{\sqrt2\pi g_A^3h_\pi^1}{2(4\pi f_\pi)^3}
\nnb \\ && \hspace*{2cm}\times
\left\{
-\frac32\left[
\frac{1}{\epsilon}
-\gamma
+{\rm ln}(4\pi)
+{\rm ln}\left(\frac{\mu^2}{m_\pi^2}\right)
+ \frac43
\right]
+3L(q)-H(q)
\right\} \, ,
\\
V_{(d)} &\!=\!&
-i(\vec{\tau}_1\times\vec{\tau}_2)^z
(\vec{\sigma}_1\!+\!\vec{\sigma}_2)\!\cdot\!\vec{q}\,
\frac{\sqrt2\pi g_A^3h_\pi^1}{2(4\pi f_\pi)^3}
\nnb \\ && \hspace*{2cm} \times
\left\{
-\frac32\left[
\frac{1}{\epsilon}
-\gamma
+{\rm ln}(4\pi)
+{\rm ln}\left(\frac{\mu^2}{m_\pi^2}\right)
+ \frac43
\right]
+3L(q)-H(q)
\right\} \, ,
\eea
where $L(q)$ and $H(q)$ are defined in Eq. (\ref{eq:lq}), 
and the Euler number is given by $\gamma=0.5772\cdots$. The quantity  
$\mu$ represents the scale of the dimensional regularization, 
and $q=|\vec{q}\,|$ with 
$\vec{q}$ defined by Eq. (\ref{eq:defmts}).
We note that we have employed, in the calculation of the diagram (d), 
the same prescription to subtract
the two-nucleon-pole contribution as in Ref.~\cite{zhu_npa05}, Eq.~(C.3) .

All these potentials from the loop diagrams 
have infinities, i.e., the $1/\epsilon$ terms. 
Along with the finite constant terms, they are renormalized by
the counter terms (PV $NN$ contact terms)
$\tilde{C}_2 +\tilde{C}_4$ and $C_6$   
(see Eq.~(5) in Ref.~\cite{zhu_npa05}).
In the maximal-subtraction scheme (MX), the $L(q)$ and $H(q)$ terms, 
which are the only ones to contribute to the finite-range potential,
are retained. This procedure fixes the associated contact term. 
In the minimal-subtraction scheme ($\overline{ \mbox{\rm MS}}$), 
besides $L(q)$ and $H(q)$ terms, an extra   
term ${\rm ln}\Big(\mu^2/m_{\pi}^2\Big)+2$ 
(rather written as $2(1\!+\!{\rm ln}(\mu/m_{\pi}))$ in the main text), 
which produces a contact term, appears.




\begin{thebibliography}{99}
\bibitem{ork_prc96}
C. Ordo\~{n}ez, L. Ray and U. van Kolck,
Phys. Rev. C \textbf{53}, 2086 (1996).

\bibitem{egm_npa00}
E. Epelbaum, W. Gl\"{o}ckle and U.-G. Mei\ss ner,
Nucl. Phys. A \textbf{671}, 295 (2000).

\bibitem{entem_prc68}
D. R. Entem and R. Machleidt, Phys. Rev. C \textbf{68}, 04001 (2003).

\bibitem{epel_05}
For a recent review, see, {\it e.g.},
E. Epelbaum, Prog. Part. Nucl. Phys. \textbf{57}, 654 (2006), and references therein.

\bibitem{weinberg}
S. Weinberg,
Phys. Lett. B \textbf{251}, 288 (1990);
Nucl. Phys. B \textbf{363}, 3 (1991).

\bibitem{zhu_npa05}
S.-L. Zhu, C. M. Maekawa, B. R. Holstein, M. J. Ramsey-Musolf
and U. van Kolck,
Nucl. Phys. A \textbf{748}, 435 (2005).

\bibitem{danilov_sjnp72}
G. S. Danilov, Sov. J. Nucl. Phys. \textbf{14}, 935 (1972).

\bibitem{missimer_prc76}
J. Missimer, Phys. Rev. C \textbf{14}, 347 (1976).

\bibitem{desplanques_npa79}
B. Desplanques and J. Missimer, Nucl. Phys. A \textbf{300}, 286 (1978). 

\bibitem{hyun_plb07} 
C. H. Hyun, S. Ando and B. Desplanques, Phys. Lett. B \textbf{651}, 257 (2007), 
nucl-th/0611018.

\bibitem{liu_xxx06} 
C.-P. Liu, Phys. Rev. C \textbf{75}, 065501 (2007).

\bibitem{pign_plb71}
D. Pignon,  Phys. Lett. \textbf{35}B, 163 (1971).

\bibitem{desp_plb72} 
B. Desplanques, Phys. Lett. \textbf{41}B, 461 (1972).

\bibitem{pirner_plb73} 
H. Pirner and D. O. Riska, Phys. Lett. \textbf{44}B, 151 (1973).

\bibitem{chem_npa74} 
M. Chemtob and B. Desplanques, Nucl. Phys. B\textbf{78}, 139 (1974).

\bibitem{kaiser_prc07}
N. Kaiser, Phys. Rev. C \textbf{76}, 047001 (2007).

\bibitem{liu_arxiv07}
Y.-R. Liu and S.-L. Zhu, nucl-th/0711.3838.

\bibitem{niskanen_arxiv07}
J. A. Niskanen, T. M. Partanen and M. J. Iqbal, nucl-th/0712.2399. 

\bibitem{desp_npa75}
B. Desplanques, Nucl. Phys. A \textbf{242}, 423 (1975).

\bibitem{hyun_chiral07}
C. H. Hyun, B. Desplanques, S. Ando and C.-P. Liu, nucl-th/0802.1606.

\bibitem{higa06}
R. Higa, M. R. Robilotta and C. A. da Rocha, 
Nucl. Phys. A \textbf{790}, 384 (2007).

\bibitem{ando06}
S. Ando and H. Fearing, Phys. Rev. D \textbf{75}, 014025 (2007).

\bibitem{amghar_npa03} 
A. Amghar, B. Desplanques and L. Theu{\ss}l, 
Nucl. Phys. A \textbf{714}, 213 (2003).

\bibitem{coon_prc86}
S. A. Coon and J. L. Friar, Phys. Rev. C \textbf{34}, 1060 (1986).

\bibitem{paris} 
M. Lacombe {\it et al.}, Phys. Rev. C \textbf{21}, 861 (1980).

\bibitem{barton_nc61}
G. Barton, Nuovo Cimento \textbf{19}, 512 (1961).

\bibitem{adler-weis}
S. L. Adler, Phys. Rev. \textbf{140}, B736 (1965), 
W. I. Weisberger, Phys. Rev. \textbf{143}, 1302 (1966).

\bibitem{DDH_ap80}
 B. Desplanques, J. F. Donoghue and B. R. Holstein,
Ann. Phys. (N.Y.) \textbf{124}, 449 (1980).

\bibitem{holstein_pr23}
B. R. Holstein, Phys. Rev. D \textbf{23}, 1618 (1981).

\bibitem{km_npa499}
N. Kaiser and U.-G. Meissner, Nucl. Phys. A \textbf{499}, 727 (1989).


\bibitem{desp_phrep98}
B. Desplanques, Phys. Rep. \textbf{297}, 1 (1998).

\bibitem{kist_prl87}
S. Kistryn \textit{et al.}, Phys. Rev. Lett. \textbf{58}, 1616 (1987). 

\bibitem{ever_plb91}
P. D. Eversheim \textit{et al.}, Phys. Lett. B \textbf{256}, 11 (1991). 

\bibitem{berdoz_prl01}
A. R. Berdoz \textit{et al.}, Phys. Rev. Lett. \textbf{87}, 272301
(2001). 

\bibitem{lansce} 
J. S. Nico and W. M. Snow, Ann. Rev. Nucl. Part. Sci. \textbf{55}, 27 (2005).

\bibitem{wir_prc95} 
R. B. Wiringa, V. G. J. Stoks and R. Schiavilla, 
   Phys. Rev. C \textbf{51}, 38 (1995).

\bibitem{sim_plb72}
M. Simonius, Phys. Lett. \textbf{41}B , 415 (1972), 
Nucl. Phys. A \textbf{220}, 269 (1974). 

\bibitem{mach_prc01}   
  R. Machleidt, Phys. Rev. C \textbf{63}, 024001 (2001).

\bibitem{stoks_prc94} 
V. G. J. Stoks {\it et al.}, Phys. Rev. C \textbf{49}, 2950 (1994).

\bibitem{amghar2_npa03} 
A. Amghar and B. Desplanques, Nucl. Phys. A \textbf{714}, 502 (2003).

\bibitem{desp_plb01}
B. Desplanques,  Phys. Lett. B \textbf{512}, 305 (2001).

\bibitem{hyun_plb01}
C. H. Hyun, T.-S. Park and D.-P. Min,  Phys. Lett. B \textbf{516}, 321 (2001).

\end{thebibliography}
\end{document}